\documentclass[fleqn,usenatbib]{mnras}

\usepackage{newtxtext,newtxmath}
\usepackage{comment}
\usepackage{longtable}
\usepackage{tablefootnote}
\usepackage{threeparttablex}
\usepackage{graphicx}
\usepackage{graphics}
\usepackage{txfonts}
\usepackage{verbatim}
\usepackage{hyperref}
\usepackage{tabularx}
\usepackage{array}
\usepackage{adjustbox}
\usepackage{pdflscape}
\usepackage[dvipsnames]{xcolor}
\newcommand{\lsun}{$\log$L/$L_{\odot}\,$}
\newcommand{\msun}{$M$/$M_{\odot}\,$}

\usepackage{multicol}
\usepackage{graphicx}
\usepackage{natbib}
\defcitealias{Desomma2020a}{DS20a}
\defcitealias{Desomma2020b}{DS20b}
\defcitealias{Caputo2000}{Caputo2000}
\defcitealias{Marconi2005}{Marconi2005}
\defcitealias{Marconi2010}{Marconi2010}

\usepackage[bottom]{footmisc}
\graphicspath{{./}{figures/}}
\usepackage[T1]{fontenc}

\DeclareRobustCommand{\VAN}[3]{#2}
\let\VANthebibliography\thebibliography
\def\thebibliography{\DeclareRobustCommand{\VAN}[3]{##3}\VANthebibliography}

\usepackage{graphicx}	
\usepackage{amsmath}	
\usepackage{amssymb}	
\usepackage{mathtools}	

\title[The Hertzsprung progression of Classical Cepheids]{The Hertzsprung progression of Classical Cepheids in the Gaia era}

\author[M. Marconi et al.]{
Marcella Marconi,$^{1}$\thanks{E-mail: marcella.marconi@inaf.it}
Giulia De Somma$^{1}$$^{2}$, Roberto Molinaro$^{1}$, Anupam Bhardwaj$^{1}$ \and Vincenzo Ripepi$^{1}$, Ilaria Musella$^{1}$, Teresa Sicignano$^{1,3}$, Erasmo Trentin$^{4}$, Silvio Leccia$^{1}$\\
$^{1}$ INAF-Osservatorio Astronomico di Capodimonte, Via Moiariello 16, 80131 Napoli, Italy\\
$^{2}$ Istituto Nazionale di Fisica Nucleare (INFN)-Sez. di Napoli, Compl. Univ.di Monte S. Angelo, Edificio G, Via Cinthia, 80126 Napoli, Italy\\
$^{3}$ Scuola Superiore Meridionale, Largo San Marcellino 10, Napoli, Italy\\
$^{4}$ Leibniz-Institut
für Astrophysik Potsdam (AIP)
An der Sternwarte 16
14482 Potsdam
}

\date{Accepted y m d. Received y m d; in original form y m d}

\pubyear{2022}

\begin{document}
\label{firstpage}
\pagerange{\pageref{firstpage}--\pageref{lastpage}}
\maketitle

\begin{abstract}
A new fine grid of nonlinear convective pulsation models for the so-called "bump Cepheids" is presented to investigate the Hertzprung progression (HP) phenomenon shown by their light and radial pulsation velocity curves. The period corresponding to the center of the HP is investigated as a function of various model assumptions, such as the efficiency of super-adiabatic convection, the mass-luminosity relation, and the metal and helium abundances. The assumed mass-luminosity relation is found to significantly affect the phenomenon but variations in the chemical composition as well as in the stellar mass (at fixed mass-luminosity relation) also play a key role in determining the value of the HP center period. Finally, the predictive capability of the presented theoretical scenario is tested against observed light curves of bump Cepheids in the ESA Gaia database, also considering the variation of the pulsation amplitudes and of the Fourier parameters $R_{21}$ and $\Phi_{21}$ with the pulsation period. A qualitative agreement between theory and observations is found for what concerns the evolution of the light curve morphology as the period moves across the HP center, as well for the pattern in period-amplitude, period-$R21$ and period-$\Phi_{21}$ planes. A larger sample of observed Cepheids with accurate light curves and metallicities is required in order to derive more quantitative conclusions.

\end{abstract}




\section{Introduction}
Classical Cepheids (hereinafter CCs) are among the most important standard candles to constrain mean and individual stellar distances and calibrate the cosmic distance scale \citep[see e.g.][and references therein]{FM10, Riess22}. At the same time, as they are intermediate-mass ($\sim$ 3 $\le$ $M$/$M_{\odot}$ $\le$ 13) stars in the central Helium-burning phase, they are excellent tracers of stellar populations with ages decreasing from hundreds of to few Myr \citep[see e.g.][]{BonoML2000, Desomma2021}. The role of preferred primary distance indicators in the calibration of the extra-galactic distance scale, through their characteristic Period-Luminosity (PL) and Period-Luminosity-Color (PLC) relations heavily relies on the structural and evolutionary properties of these stars. First, at fixed chemical composition, the period of oscillation is well known to be related to the mean stellar density, i.e. to the mass, the luminosity, and the effective temperature of the star. This relation implies the existence of a PLC relation because the stellar mass and luminosity are related to each other as predicted by stellar evolution  (Mass-Luminosity, hereinafter ML relation) for intermediate-mass stars in the central Helium burning phase (blue loop phase in the Color-Magnitude diagram). The PL relation is then statistically obtained by averaging over the color extension of the instability strip. The reddening-free formulation of the PL relation, i.e. the Period-Wesenheit (PW) relation, partially corrects for the finite width of the instability strip, by introducing a color-term whose coefficient is fixed as the ratio between the total and the selective extinctions in the chosen photometric bands \citep[since][]{Madore1982}. 
Thus, the ML relation of CCs plays a relevant role in determining the coefficients of the relations that make these pulsators distance indicators. Several empirical and theoretical investigations, involving e.g. CCs in eclipsing binary systems \citep[see e.g.][]{Pietr2010, Pietr2011} or the model fitting of observed light and radial velocity curves \citep[see e.g.][and references therein]{Keller2006, Marconi2013a, Marconi2013b, Marconi2017, Ragosta2019} or the application of mass-dependent relations to target CCs with known distances \citep[see e.g.][and references therein]{Caputo2005, Marconi2020}, have demonstrated that the ML relation is likely marginally dispersed and brighter than the relation expected when neglecting mass-loss, core overshooting and rotation. These results suggest that some combinations of these so-called non-canonical phenomena are expected to be at work in the structures of CCs.\par
Several of the adopted methods to derive CC properties and distance scale involve the investigation of the morphology of light and radial pulsation velocity curves that are known to depend on the input physical and chemical parameters. Indeed, the shape and the amplitude of the luminosity and radial velocity variations depend on the position within the instability strip, as well as on the adopted metallicity and helium abundance \citep[see e.g.][and references therein]{Bonocurve2000, Fiorentino2002, Marconi2005}. A peculiar property of CC light and radial velocity curves is the so-called Hertzsprung progression (HP). This phenomenon was discovered about one century ago \citep{Hertzprung1926} when, investigating a sub-sample of Galactic CCs, Hertzsprung found a relationship between the position of the secondary maximum, called bump, along the light curve and the pulsation period. A similar relation was also detected among Magellanic Clouds (MCs) and Andromeda CCs \citep{Payne1947, Payne1954, Shapley1940}. Moreover, the same relation was discovered in radial velocity curves by \citet{Joy1937, Ledoux1958}. The period values affected by this phenomenon range from $\sim$ 6 to $\sim$ 16 d, and the signature is the appearance of a bump along both the light and the radial velocity curves so that these pulsators are called {\it bump} Cepheids. This secondary feature is observed on the descending branch of the light and radial velocity curves for CC periods up to 9 days, close to the main light/radial velocity maximum for periods ranging from $\sim$  9 to $\sim$ 12 d, and at earlier phases for longer periods. The origin of the HP has been widely debated in the literature. \citet{Simon1981} found that both the phase difference $\Phi_{21}$ and the amplitude ratio $R_{21}$ show a sharp minimum close to the HP center. Subsequently, \citet{Moskalik1992, Moskalik2000} suggested that the minimum in the Fourier parameters for Galactic Cepheids corresponded to $P_{HP} \sim 10.0$ d, while \citet{Welch1995} found $P_{HP}$ = $11.2 \pm 0.8$ d investigating a large sample of Large Magellanic Cloud (LMC) CCs. This result supported the shift of the HP center toward longer periods moving from the Milky Way to the LMC originally suggested by \citet{Payne1947} but also claimed by \citet{Andreasen1987} and \citet{Andreasen1988}. Subsequently, \citet{Beaulieu1998} suggested that the HP center for LMC and  Small Magellanic Cloud (SMC) CCs corresponded to $P_{HP}$ = $10.5 \pm 0.5$ d and $P_{HP}$ = $11.0 \pm 0.5$ d, respectively. This result confirmed that a decrease in metallicity moves the HP center toward longer periods.
More recently, \citet{bhard15} showed that the central period of the HP increases with wavelength in the case of the Fourier amplitude parameters and decreases with increasing wavelength in the case of phase parameters.
The central minimum of the HP for amplitude parameters was also found to shift to longer periods with a decrease/increase in metallicity/wavelength for both theoretical and observed light curves \citep{bhard15,bhard17}.
The main possible scenarios, quoted in the literature, to explain the HP phenomenon are:
\begin{enumerate}[i)]
\item the echo model first proposed by \citet{Whitney1956} and subsequently discussed by \citet{Christy1968, Christy1975} based on CC non-linear, radiative models. These models suggested that, during each pulsation cycle, at the phases of minimum radius just before the maximum expansion velocity a pressure excess was produced in the first He ionization region. The resulting rapid expansion was able to generate an outward and an inward pressure wave. The latter was predicted to reach the stellar core, near the phase of maximum radius, to reflect and then reach the surface one cycle later, producing the bump;
\item the resonance model suggested by \citet{Simon1976} based on linear, adiabatic models that predicted a resonance between the second overtone (SO) and the F mode when their period ratio is close to 0.5. Indeed the F mode instability was predicted to drive the SO instability through a resonance mechanism.
\end{enumerate}

The echo hypothesis was analysed in detail in a series of papers by \citet{W83,AW84,AW85}. These authors concluded that the temporal resonance condition for the inward pulse can be satisfied near the resonance between the second overtone and the fundamental period \citep[see][for details]{AW84}. Moreover, based on the calculated acoustic fluxes, they suggested that the  mode-resonance model is more appropriate than the pulse-resonance model for bump Cepheids.
 On the theoretical side, \citet{BonoHP2000} presented the results of an extensive theoretical investigation on the pulsation behavior of Bump CCs adopting a chemical composition typical of LMC CCs (Y=0.25, Z=0.008), stellar masses ranging from 6.55 to 7.45 $M_{\odot}$ and a canonical ML relation \citep{Castellani1992}. The results of these computations showed the HP progression, in the sense that as the models move from the blue to the red edge of the IS  the bump is at first located along the descending branch, then it crosses the luminosity/velocity maximum and subsequently it appears along the rising branch. The predicted period at the HP center was found to be $P_{HP}$ = $11.24 \pm 0.46$ d in very good agreement with the empirical value based on the previous analysis of Fourier parameters of LMC CC light curves \citep[P\textsubscript{HP}= 11.2 $\pm$ 0.8 d,][]{Welch1997}. Moreover, \citet{BonoHP2000} found that both the skewness and the acuteness of the predicted light curves typically show a well-defined minimum at the HP center in good agreement with empirical estimates. For the models at the HP center, the period ratio between the second overtone and the fundamental (F) mode was predicted to roughly range from 0.51 (red models) to 0.52 (blue models). In this paper, we extend the analysis to other chemical compositions, with metallicities Z ranging from 0.004 to 0.03, also increasing the helium abundance Y at fixed solar metallicity, and investigate for the first time the effect of the ML relation and the efficiency of superadiabatic convection on the predicted properties of Bump CCs. Moreover, we compute a much finer grid  (for effective temperatures, mass and luminosity) than performed in our previous investigations \citep[see e.g.][and references therein]{Desomma2022, Desomma2021, Desomma2020b, Desomma2020a, Marconi2005, Marconi2010, Marconi2020} to provide an exhaustive theoretical scenario for the comparison with observations, in particular with Gaia in the Milky Way.

The organization of the paper is as follows. In Section 2, we present the computed theoretical scenario. The resulting HP phenomenon for light and radial velocity curves as a function of chemical composition, ML relation, and super-adiabatic convection, is discussed in Section 3, while Section 4 is devoted to the comparison with the observations. The Conclusions and some final remarks close the paper.

\section{The theoretical framework}
In this section, we present the new computed models adopted to investigate the HP phenomenon. The physical and numerical assumptions at the basis of these nonlinear hydrodynamic computations are the same as in our previous papers \citep[see e.g.][and references therein]{BonoHP2000, Marconi2005}. Five chemical compositions have been taken into account, namely  $Z$=$0.004$ $Y$=$0.25$, $Z$=$0.008$ $Y$=$0.25$, $Z$=$0.02$ $Y$=$0.28$, $Z$=$0.03$ $Y$=$0.28$ and $Z$=$0.02$ $Y$=$0.30$. In order to cover the parameter space of Bump CCs, for each chemical composition, we limited the stellar mass to the range from 5 to 8 $M_{\odot}$ with a step of 0.2 $M_{\odot}$. For each stellar mass, we adopted two luminosity levels, corresponding to a canonical \citep[as given in][]{BonoML2000} and a moderate noncanonical ML relation (increasing the canonical ML by 0.2 dex,  \citep[see][]{Desomma2020b, Desomma2022}, respectively. The effective temperature was varied from 3900 to 7100 K to evaluate the location of the F boundaries for each combination of M, L, $Z$, and $Y$. The adopted effective temperature step, 50 K, allowed us to evaluate the F boundaries with an error of $\pm$25 K and, at the same time, to explore in detail the evolution of the bump phase position across the instability strip. The adopted mass and luminosity values, with the indication of the corresponding ML relation (A for the canonical and B for the noncanonical case) for each assumed chemical composition, are reported in Table \ref{param_HP_models}. To investigate the effect of a variation in the efficiency of super-adiabatic convection, two different values of the mixing length parameter $\alpha_{ml}$, defined as the ratio between the mixing length and the pressure height scale, are assumed, as reported in the fifth column of Table \ref{param_HP_models}. We notice that the convective treatment adopted in the employed pulsation hydro-code is not the standard mixing length theory but a mixing length parameter is assumed to close the non-linear system of dynamical and convective equations \citep[see e.g.][for details]{bs94}. The last two columns of the same Table list the effective temperatures corresponding to the location of the F blue and red instability strip edges hereinafter indicated as FBE and FRE, respectively. As the effective temperature step in the model computation was 50 K, the effective temperature hotter/cooler by 25 K than the bluest/reddest pulsating models was adopted as the FBE/FRE location.
We notice that First Overtone models are not computed in this work as the HP phenomenon is observed in Fundamental mode Cepheids.

\begin{figure*}
\centering
\includegraphics[width=1.0\textwidth]{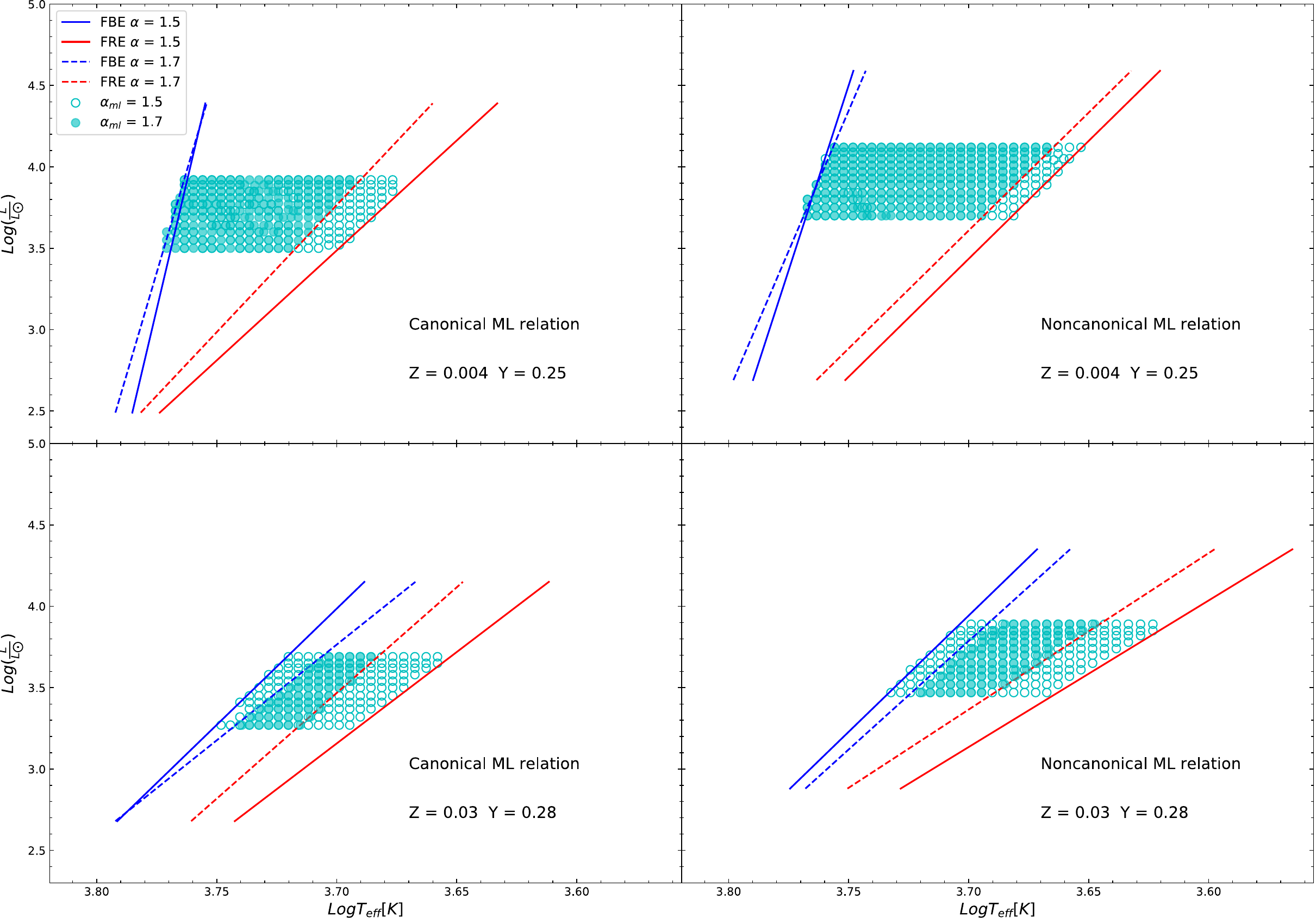}
\caption{Distribution of computed models (cyan symbols) in the HR diagram as compared with the predicted fundamental instability strip boundaries \citep[from][]{Desomma2020a, Desomma2022} both for $\alpha_{ml}$=$1.5$ (solid lines) and $\alpha_{ml}$=$1.7$ (dashed lines) for the lowest (upper panels) and highest (lower panels) considered metal abundance.}
\label{Fig:strip_models_smc_m31}
\end{figure*}

As an example, the left panel of Fig. \ref{Fig:strip_models_smc_m31} shows the location of the computed canonical models for the labeled solar chemical compositions, over-imposed to the corresponding complete instability strip predicted in \citet[][]{Desomma2020a, Desomma2022}. Open symbols correspond to $\alpha_{ml}$ = $1.5$, whereas filled symbols show the location of $\alpha_{ml}$ = $1.7$ models. The right panel of Fig. \ref{Fig:strip_models_smc_m31} shows the same kind of comparison for non-canonical models.

\section{The Hertzprung progression of Classical Cepheids from the light and pulsation velocity curves}

\begin{figure}
\centering
\includegraphics[trim={0cm 5.0cm 2.0cm 6cm},clip, width=0.5\textwidth ]{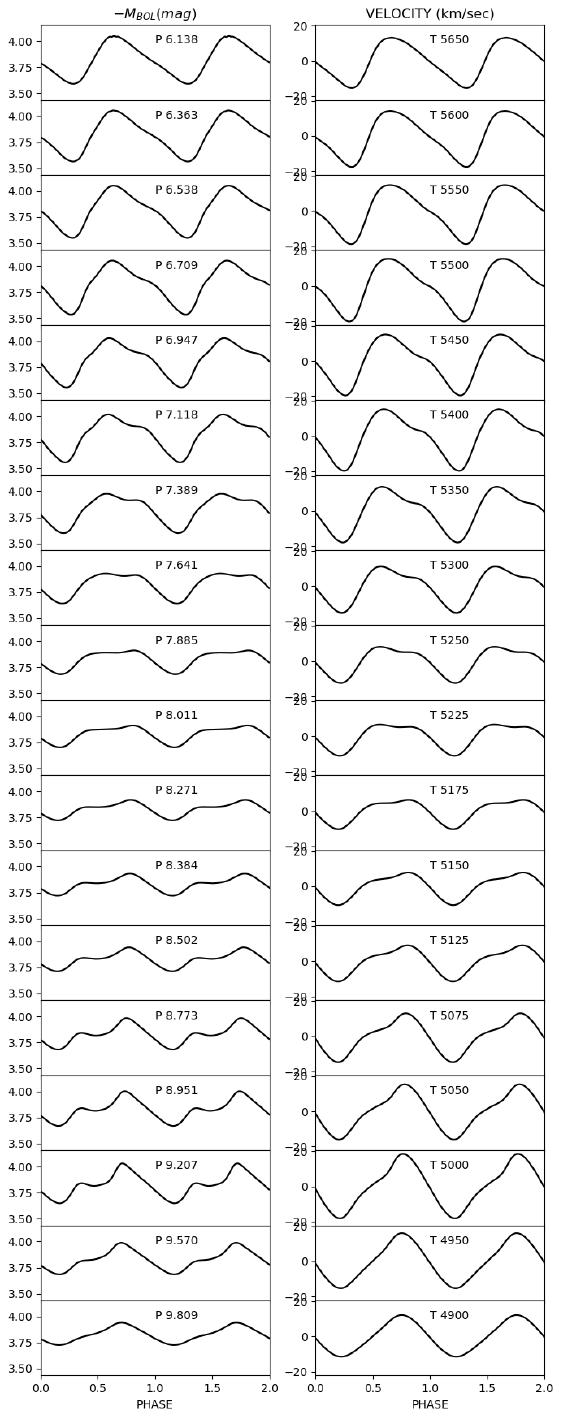}
\caption{Bolometric light curves (left panel) and pulsation velocity curves (right panel) for a sequence of nonlinear canonical models derived for $Z$ = $0.02$, $Y$ = $0.28$, $M$=$6.2M_{\odot}$ and $\alpha_{ml}$=$1.5$.}
\label{Fig:prog_mw_6p2}
\end{figure}

One of the main outputs of the adopted nonlinear convective hydrodynamic models is the prediction of the variations of all the relevant stellar quantities along the pulsation cycle.
Figure \ref{Fig:prog_mw_6p2} shows an example of the predicted bolometric light (left panels) and radial pulsation velocity (right panels) curves for $M$=6.2$M_{\odot}$, canonical ML relation and $\alpha_{ml}$=$1.5$ at fixed chemical composition $Z$=$0.02$, $Y$=$0.28$. The entire atlas of light and radial velocity curves for all the investigated chemical compositions and masses, for both the assumed ML relations, varying the efficiency of super-adiabatic convection, is available as supplementary material. The pulsation period (left panels) and the effective temperature (right panels) are labeled for each pulsation model.
We notice that, as the pulsation period increases, the morphology of the predicted curves changes, with the {\it bump} moving in phase from the descending to the rising branch, across models that show quite flattened curves, showing a bump equivalent in brightness to the curve true maximum. These transition models correspond to the so-called HP center and cover slightly different parameters when switching from light to pulsation velocity curves. In general, the HP center is anticipated by 100-200 K in the bolometric/optical band light curves with respect to the pulsation velocity ones. In most cases, the HP center also corresponds to a local minimum in the pulsation amplitudes.

\clearpage

\begin{figure*}
\centering
\includegraphics[width=1.0\textwidth]{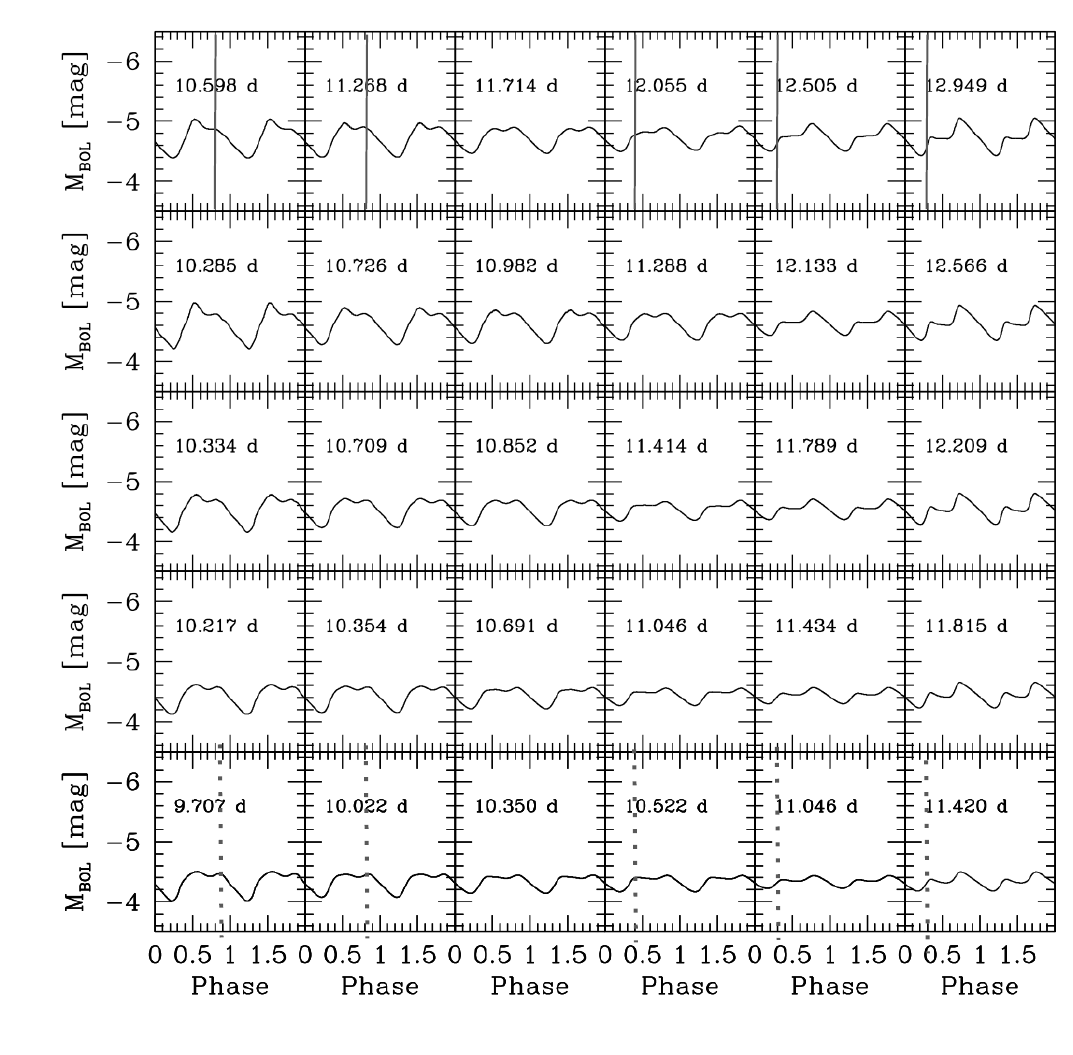}
\caption{Model bolometric light curves for  $Z$=$0.008$, $\alpha_{ml}$  = $1.5$ and canonical ML relation, across the HP center, increasing the stellar mass from 6.8 $M_{\odot}$ (bottom panels)  to 7.6 $M_{\odot}$ (top panels). The model period value (increasing from left to right) is labeled in each panel. The vertical dotted and solid lines mark the position of the bump before and after the HP center for the 6.8 $M_{\odot}$ and 7.6 $M_{\odot}$ model curves, respectively.}
\label{Fig:figHPM}
\end{figure*}

\begin{figure*}
\centering
\includegraphics[width=1.0\textwidth]{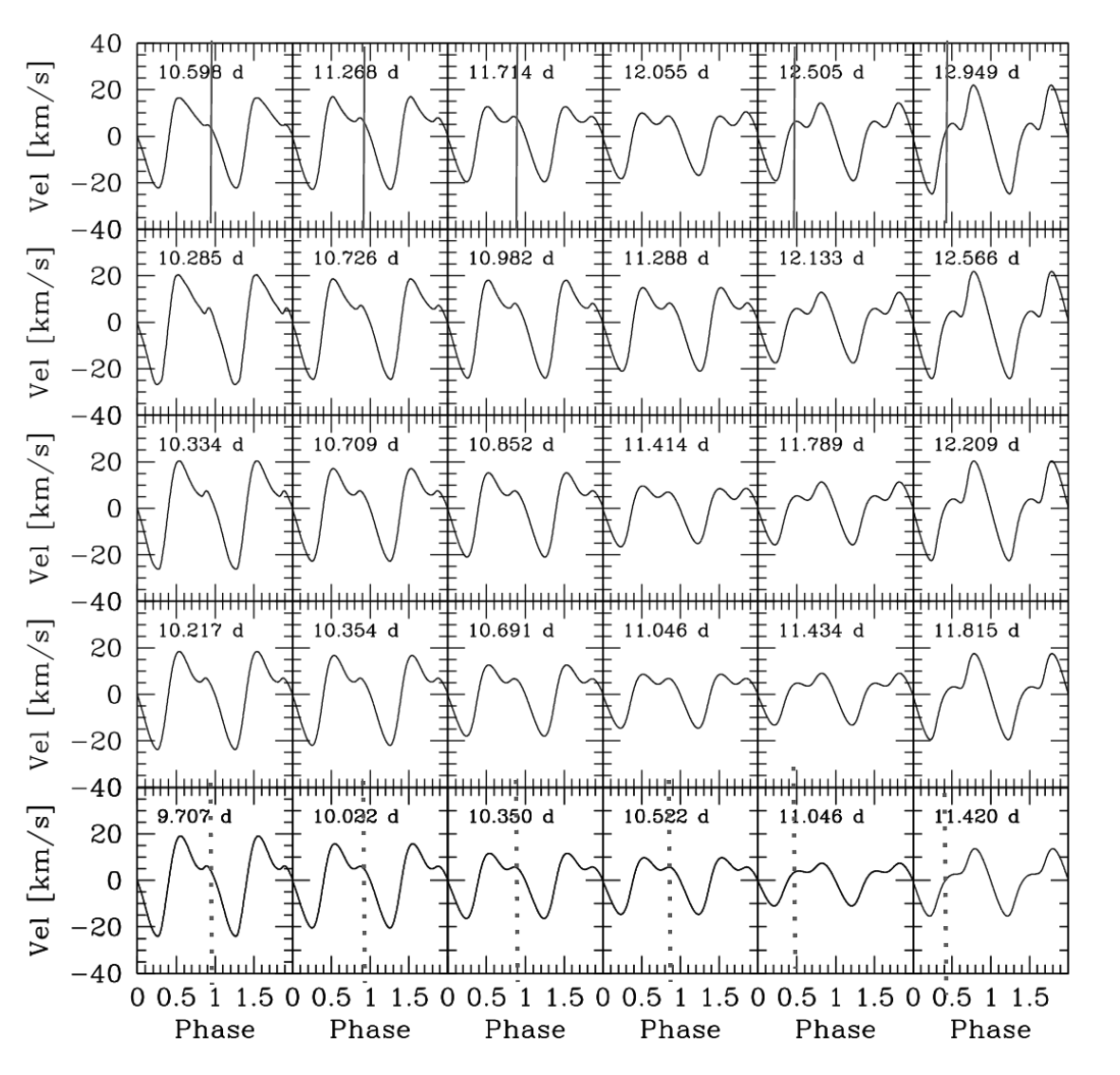}
\caption{The same as in Fig. \ref{Fig:figHPM} but for radial velocity curves.}
\label{Fig:figHPV}
\end{figure*}

In order to investigate in more detail the behaviour of model curves close to the HP center, in Figures \ref{Fig:figHPM} and \ref{Fig:figHPV}, we show the bolometric light and the pulsation velocity curves, respectively, for the 5 central models (effective temperature/period decreasing/increasing from left to right) of the computed sets for $Z$=$0.008$ and canonical ML relation, at fixed efficiency of super-adiabatic convection ($\alpha_{ml}$=$1.5$), increasing the stellar mass from 6.8 $M_{\odot}$ (bottom panels)  to 7.6 $M_{\odot}$ (top panels) with a step of 0.2 $M_{\odot}$.
The vertical dotted and solid lines mark the position of the bump before and after the HP center for the 6.8 $M_{\odot}$ and 7.6 $M_{\odot}$ model curves, respectively. 
Inspection of the bolometric light curves suggests that an increase of the model mass tends to move the center of the HP towards slightly longer periods, namely from around 10.35 d for 6.8$M_{\odot}$ to around 11.7 d for 7.6$M_{\odot}$.
A similar trend is observed for the pulsation velocity curves, with the HP center period changing from something between 10.5 and 11.0 d for 6.8$M_{\odot}$ to around 12.0 d for 7.6$M_{\odot}$.

\begin{figure*}
\centering
\includegraphics[width=1.0\textwidth]{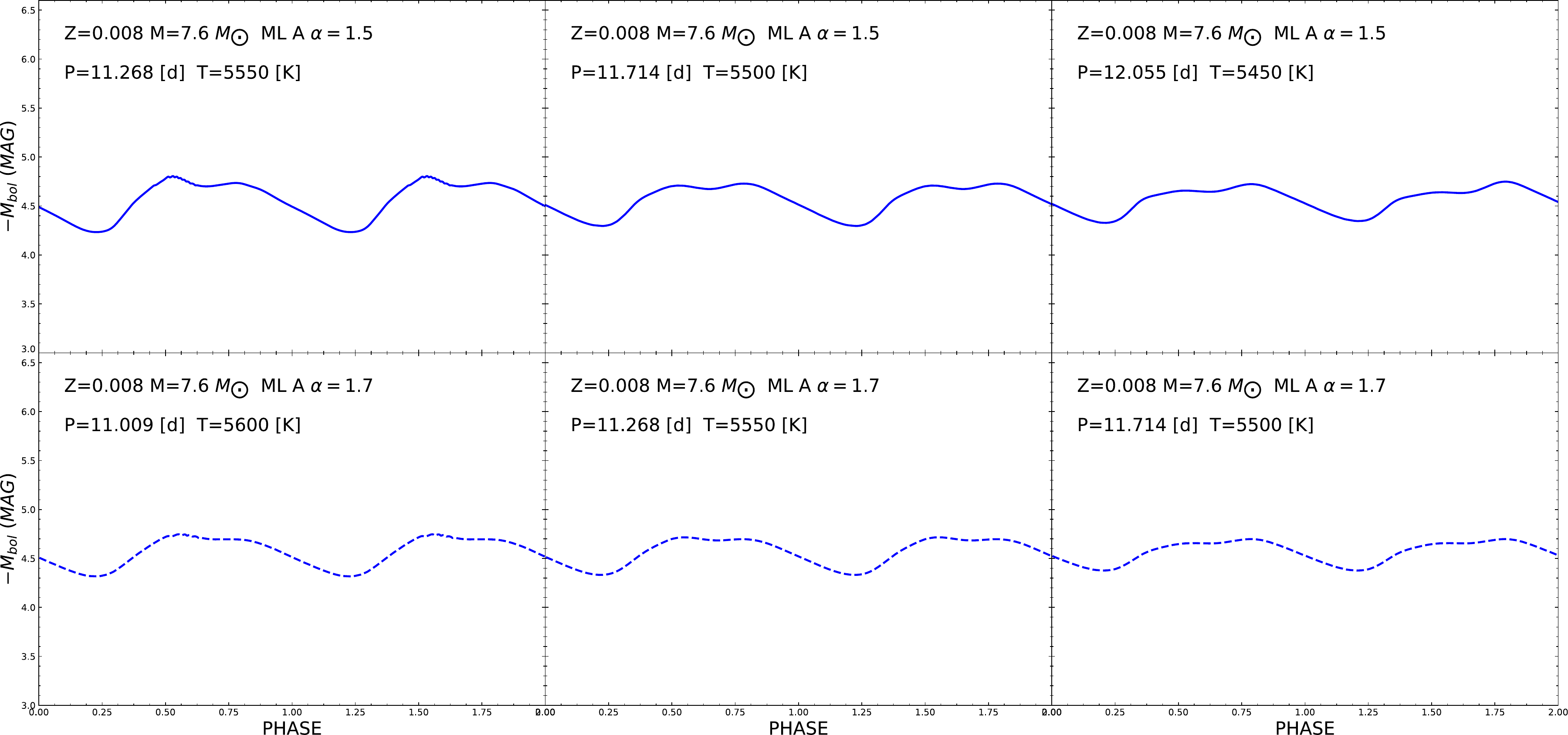}
\caption{Model bolometric light curves for $\alpha_{ml}$  = $1.5$ (upper panels) and $\alpha_{ml}$ = $1.7$ (lower panels) across the HP center, for the labelled metallicity, period and effective temperature, canonical ML relation (case A) and $M$=$7.6M_{\odot}$.}
\label{Fig:lc_lmc_alfaeff_7p6}
\end{figure*}

\begin{figure*}
\centering
\includegraphics[width=1.0\textwidth]{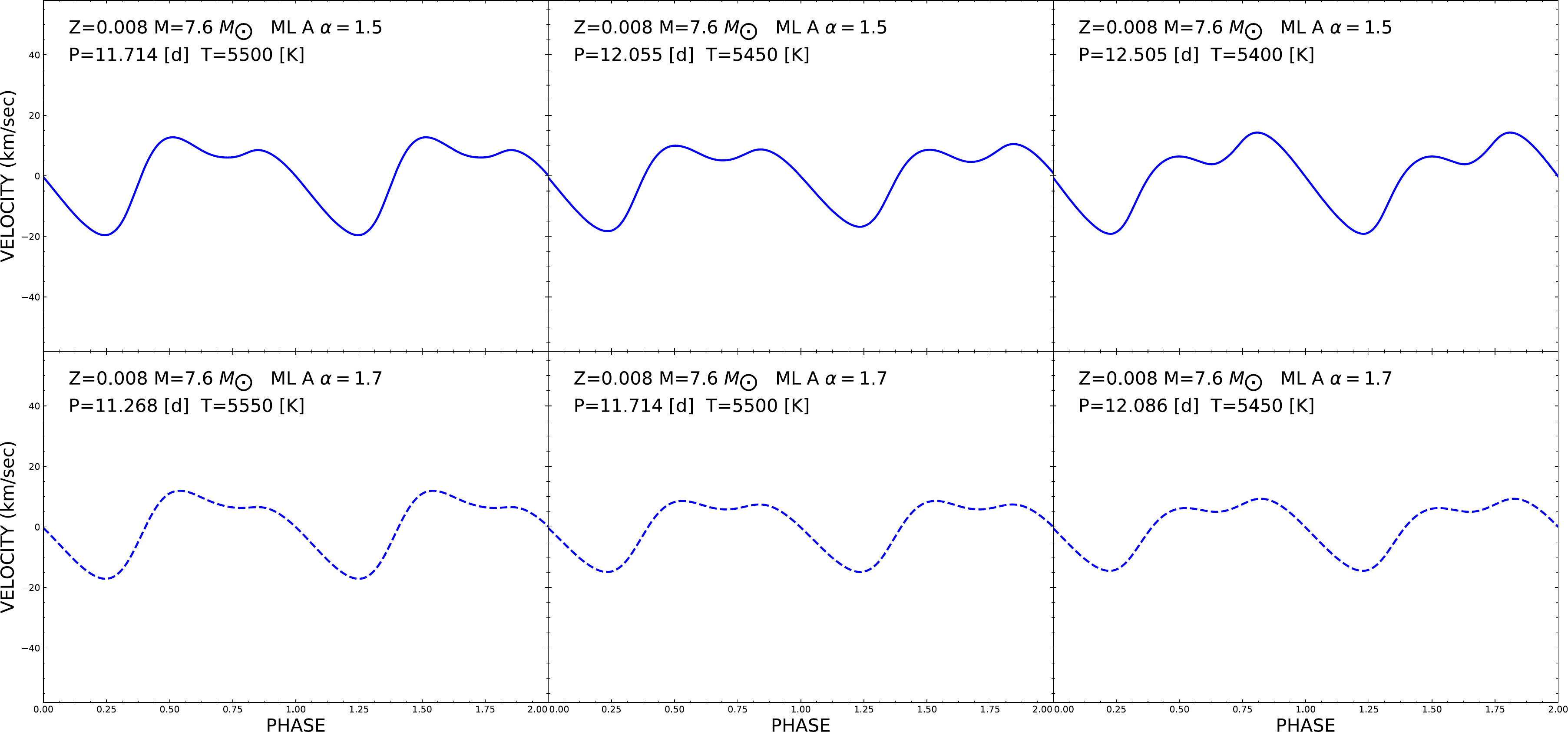}
\caption{The same as in Fig. \ref{Fig:lc_lmc_alfaeff_7p6} but for radial velocity curves.}
\label{Fig:vc_lmc_alfaeff_7p6}
\end{figure*}

In order to investigate the effect of our assumption on the efficiency of super-adiabatic convection, in Figures \ref{Fig:lc_lmc_alfaeff_7p6} and \ref{Fig:vc_lmc_alfaeff_7p6} we show the bolometric light and the radial velocity curves for the three central models of the computed sets for $Z$=$0.008$ $M$=$7.6M_{\odot}$ and canonical ML relation, varying the efficiency of super-adiabatic convection from $\alpha_{ml}$=$1.5$ (upper panels of both figures) to $\alpha_{ml}$=$1.7$  (lower panels of both figures). 
We notice that a variation in the efficiency of convection slightly modifies the amplitude of the curves but with no significant effect on the morphology and in turn on the HP center period and temperature. This is also connected to the insensitivity of the second overtone - fundamental mode (hereinafter P2/P0) resonance to variations of the mixing length parameter.

\begin{figure*}
\centering
\includegraphics[width=1.0\textwidth]{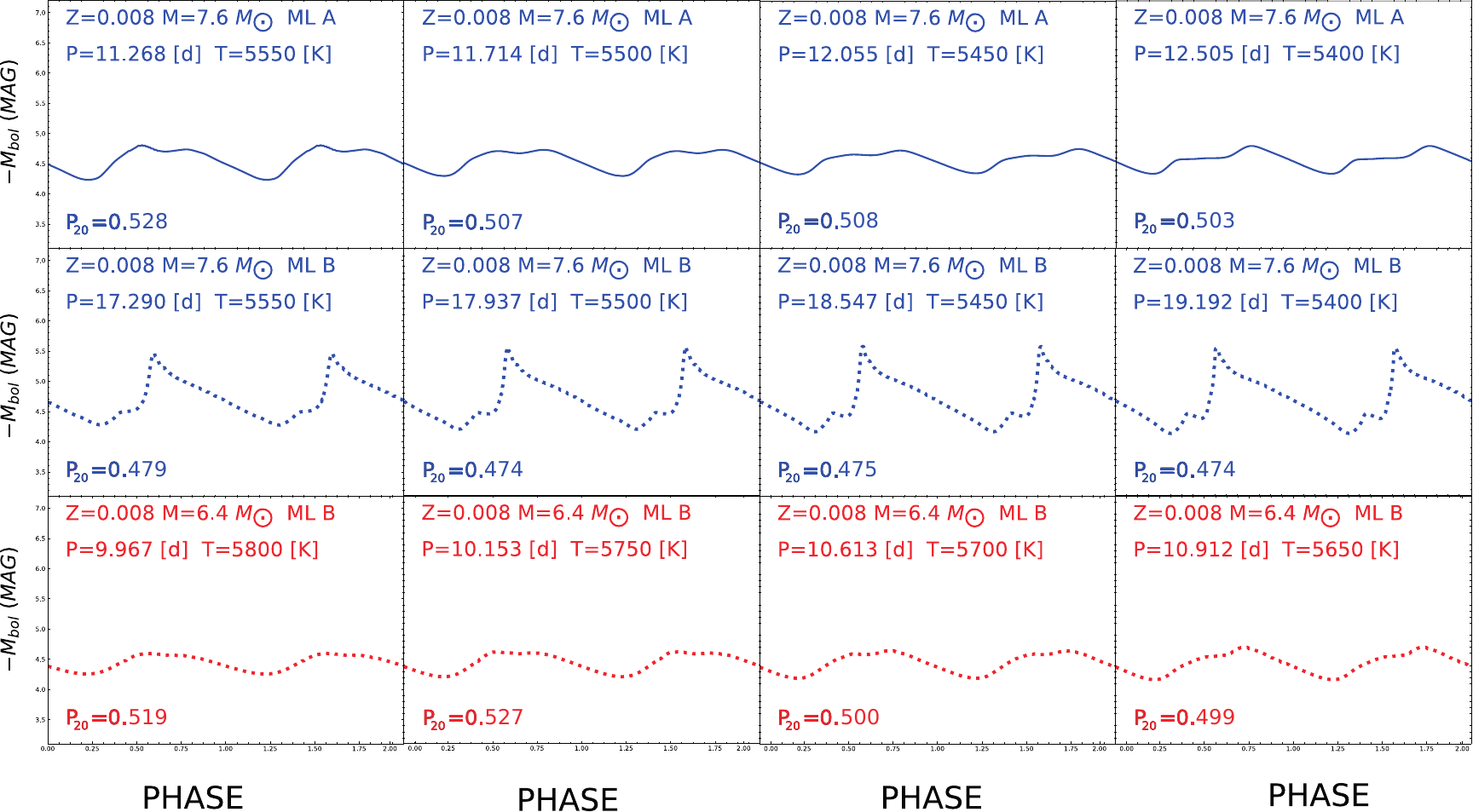}
\caption{Model bolometric light curves assuming a canonical (upper panels) and a noncanonical (middle panels) ML relation, across the HP center, for the labelled metallicity, period and effective temperature, $\alpha_{ml}$=$1.5$ and $M$=$7.6M_{\odot}$. The lower panels show noncanonical model light curves across the HP center for $M$=$6.4M_{\odot}$. {\ bf The labelled P20 is the ratio between the second overtone and fundamental mode periods (P2/P0).}}
\label{Fig:lc_lmc_mleff}
\end{figure*}

\begin{figure*}
\centering
\includegraphics[width=1.0\textwidth]{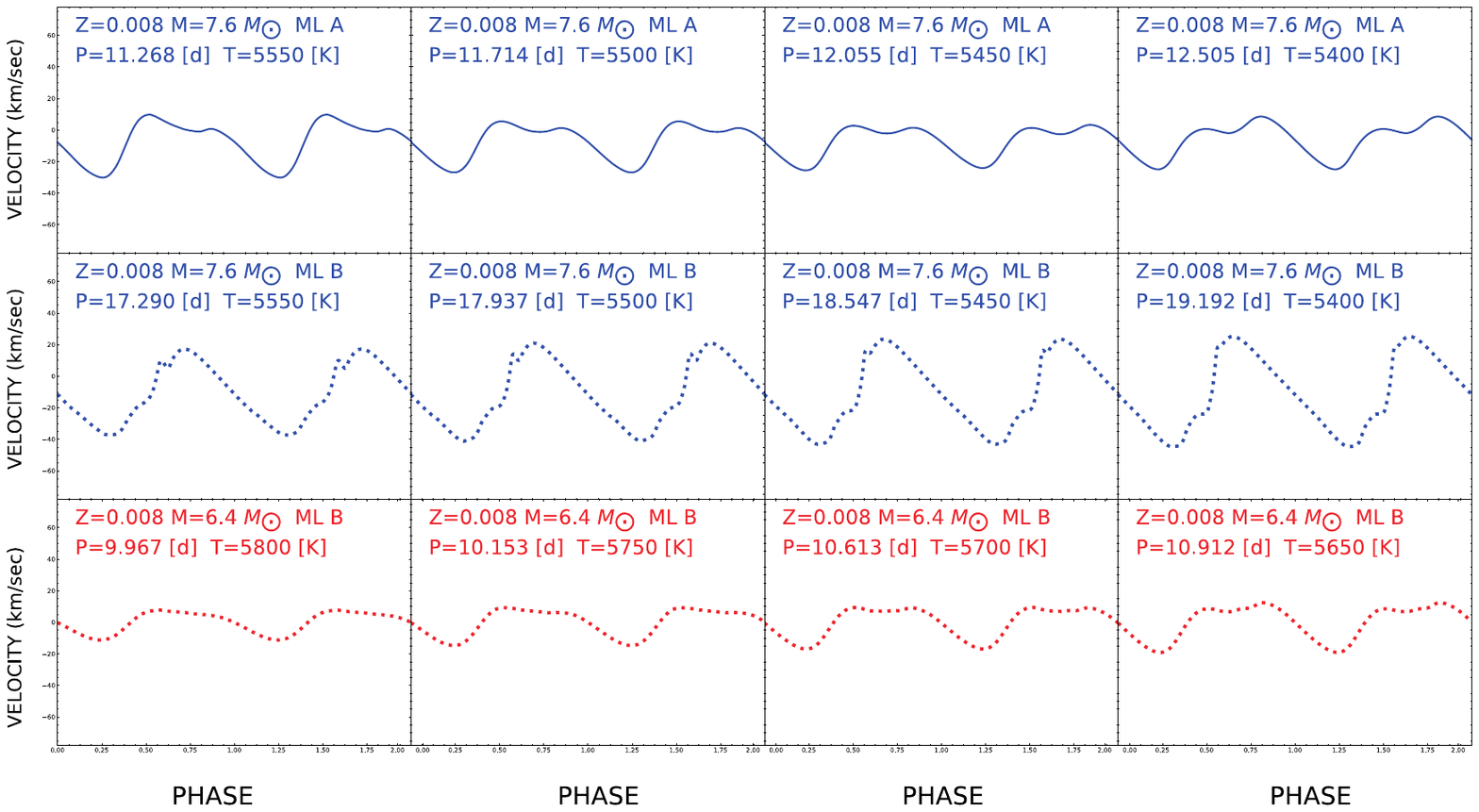}
\caption{The same as in Fig. \ref{Fig:lc_lmc_mleff} but for radial velocity curves.}
\label{Fig:vc_lmc_mleff}
\end{figure*}

As for the effect of the ML relations, Figures \ref{Fig:lc_lmc_mleff} and \ref{Fig:vc_lmc_mleff} show the same sets of models close to the HP center but varying the ML relation from canonical (case A, upper panels) to moderately noncanonical (case B, middle panels).
We notice that the amplitude and the morphology of the curves are completely modified when the ML relation is changed as an effect of the significant period and density change.
Indeed the HP center is obtained for a significantly lower mass and different effective temperatures, as shown in the lower panels of both figures (see labelled parameters).
We notice that both the period and the stellar properties of models depicting the HP center, as well as the P2/P0 resonance center, depend on the assumed ML relation. As a consequence, at fixed chemical composition, the measured period at the HP center could provide useful constraints on the ML relation of observed bump Cepheids, even if this dependence is expected to be at least in part degenerate with the above-discussed dependence on the stellar mass, at fixed ML relation (see Figures \ref{Fig:figHPM} and \ref{Fig:figHPV}). Indeed the HP center period is expected to increase as the ML relation gets fainter and as the stellar mass increases at fixed ML relation. 
 The labelled P2/P0 values (P20) confirm that only decreasing the stellar mass noncanonical models show the HP center, at fixed metallicity. Indeed P20 remains close to 0.5 only for models in the upper and bottom panels, attaining smaller values for models in the middle panels.
  The evidence that models with similar P2/P0 have similar light and velocity curves seems to support the crucial role of resonance in shaping the Hertzsprung Progression.

\subsection{The effect of the adopted chemical composition}

\begin{figure*}
\centering
\includegraphics[width=1.0\textwidth]{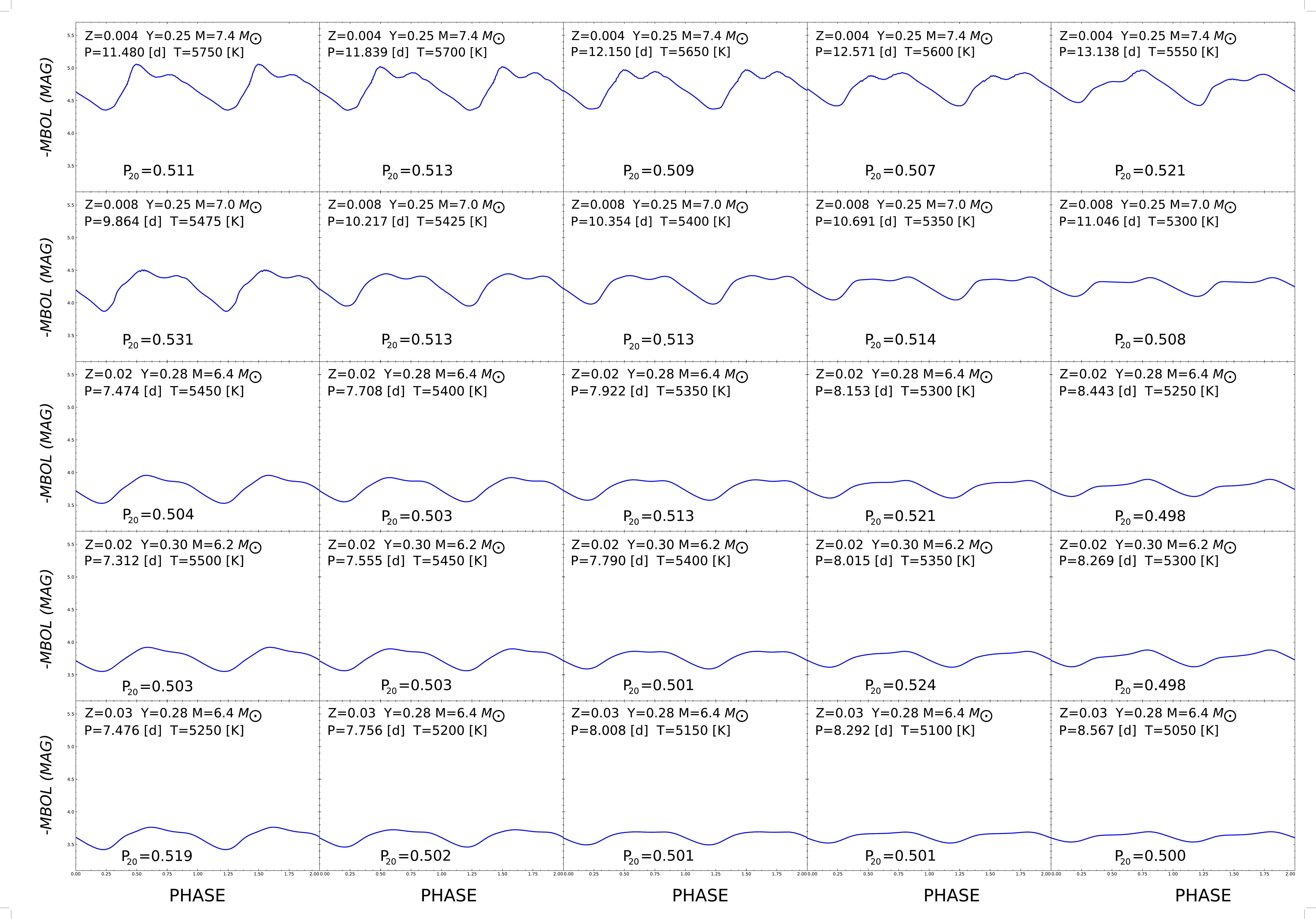}
\caption{Model bolometric light curves with $\alpha_{ml}$  = $1.5$ and canonical ML relation (case A), across the HP center, by increasing the metallicity from $Z$=$0.008$ up to $Z$=$0.03$ (see labels). In each panel, the period in days and the effective temperature in kelvin are labeled. P20, as labeled, denotes the ratio of periods between the second overtone and the fundamental mode.}
\label{Fig:lc_z_eff}
\end{figure*}

\begin{figure*}
\centering
\includegraphics[width=1.0\textwidth]{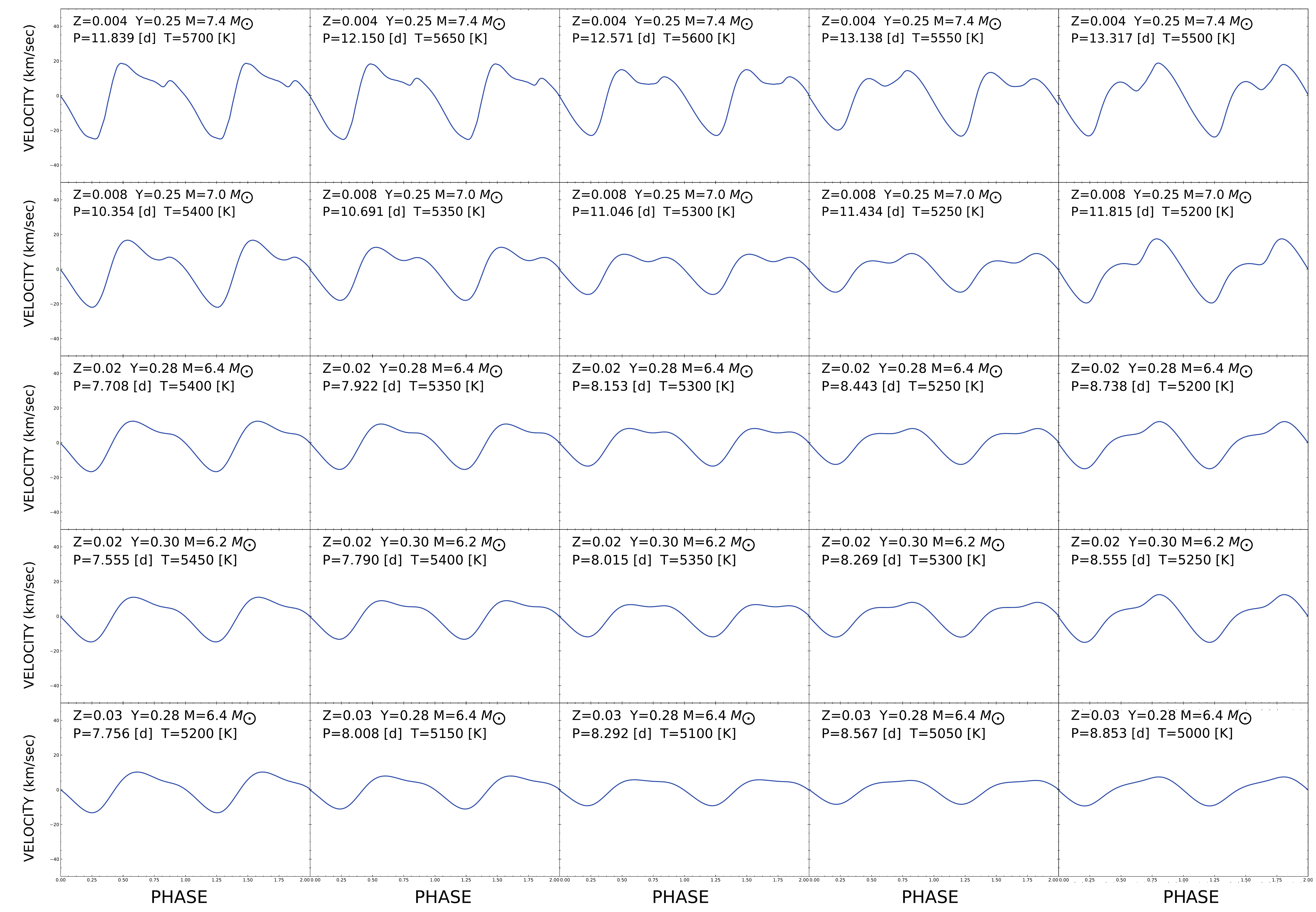}
\caption{The same as in Fig. \ref{Fig:lc_z_eff} but for radial velocity curves.}
\label{Fig:vc_z_eff}
\end{figure*}

The same analysis performed for $Z$=$0.02$, $Y$=$0.28$, can be repeated for all the adopted chemical compositions. Figures \ref{Fig:lc_z_eff} and \ref{Fig:vc_z_eff} show the behaviour of the predicted bolometric light and pulsation velocity curves, respectively, for selected M, L, $T_e$ combinations that allow the center of the HP   (where the bump gets the closest to the maximum) to occur in the middle of the instability strip. We notice that, both in the case of the light curves and in the case of pulsation velocity variations, the period corresponding to the center of the HP moves towards longer values as the metal abundance decreases, with an effect that is more important than the dependence on the adopted stellar mass, changing from $\sim$ 8.0 d to $\sim$ 11.4 d in the case of the light curves and from $\sim$ 8.3 d to $>$ 11.6 d in the case of the radial pulsation velocity curves when the metallicity decreases from $Z$=$0.03$ to $Z$=$0.004$. 
No significant effect appears to occur when moving from $Z$=$0.02$ to $Z$=$0.03$ or varying $Y$ from $0.28$ to $0.30$ at fixed $Z$=$0.02$.
 According the plotted P2/P0 (P20) values, we notice a small sensitivity of the P2/P0 resonance center to metallicity.

\begin{figure*}
\centering
\includegraphics[width=1.0\textwidth]{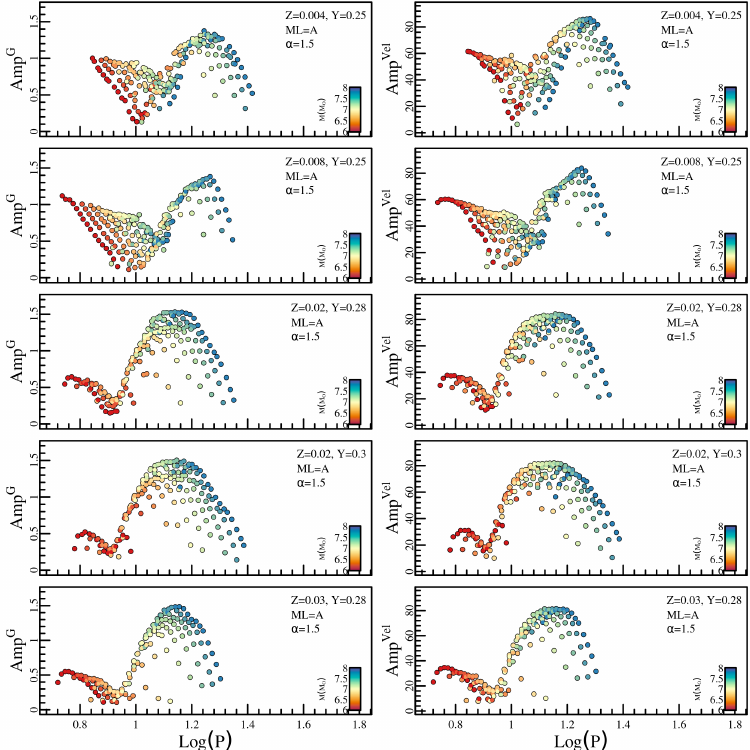}
\caption{The variation of the predicted Gaia G-band (left panels) and radial velocity (right panels) amplitudes as a function of the pulsation period for canonical models (case A) with the labelled chemical compositions. The mass value is color-coded in each panel.}
\label{Fig:amp_can}
\end{figure*}

\begin{figure*}
\centering
\includegraphics[width=1.0\textwidth]{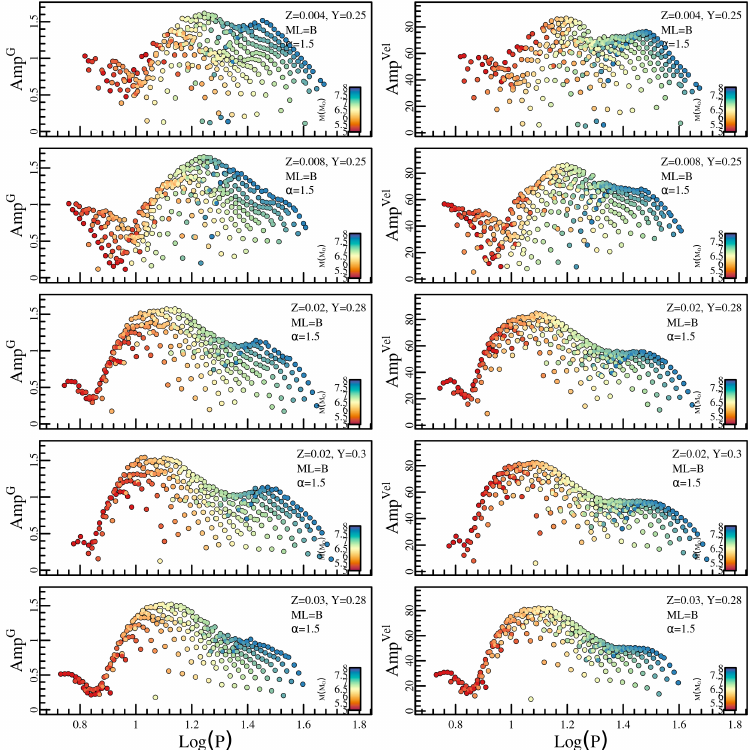}
\caption{The same as in Fig. \ref{Fig:amp_can} but for moderately noncanonical models (case B).}
\label{Fig:amp_ncan}
\end{figure*}

The bolometric light curves of all the computed models have been transformed into the Gaia filters by adopting PHOENIX model atmospheres \citep[][]{Chen2019}, so that mean magnitudes, colors, and amplitudes in the three Gaia filters could be derived.
Figures \ref{Fig:amp_can} and \ref{Fig:amp_ncan} show the predicted Gaia G-band light curve (left panels) and radial pulsation velocity (panels) amplitudes as a function of the pulsation period for canonical (case A) and noncanonical (case B) model sets, respectively. The adopted chemical composition is labelled and the stellar mass is color-coded in each panel.

Inspection of these plots suggests that in the case of canonical models (see Fig. \ref{Fig:amp_can}) a local minimum amplitude is found for a period value that corresponds to the HP center and varies not only with the mass and the assumed ML relation but also with the assumed chemical composition, specifically moderately increasing as the metallicity decreases. In the case of noncanonical models (see Fig. \ref{Fig:amp_ncan}) we observe the same trend with metallicity with the minimum shifting toward a shorter period and becoming better defined as $Z$ increases from $0.004$ to $0.03$.

\begin{figure*}
\centering
\includegraphics[width=1.0\textwidth]{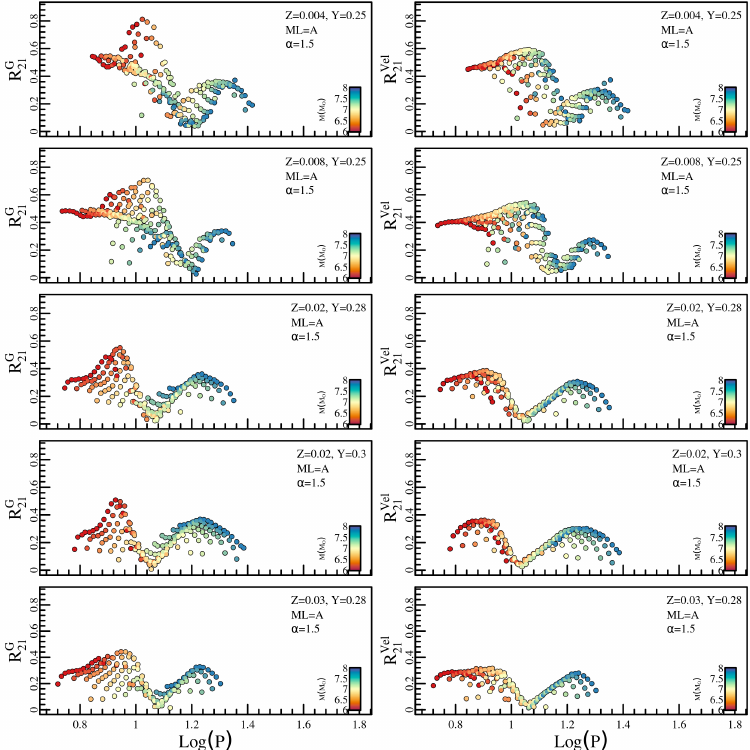}
\caption{The variation of the theoretical Fourier parameter $R_{21}$ as derived from the G band light curves (left panels) and the radial pulsation velocity curves (right panels) of canonical models (case A), as a function of the pulsation period for the labelled chemical composition. The adopted stellar mass is color-coded.}
\label{Fig:R21_can}
\end{figure*}

\begin{figure*}
\centering
\includegraphics[width=1.0\textwidth]{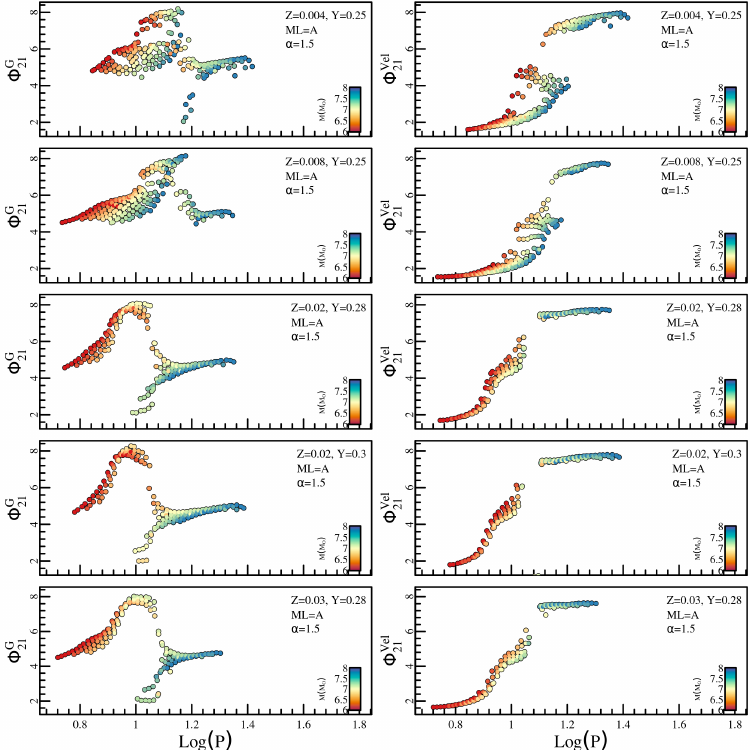}
\caption{The variation of the theoretical Fourier parameter $\Phi_{21}$ as derived from the G band light curves (left panels) and the radial pulsation velocity curves (right panels) of canonical models (case A), as a function of the pulsation period for the labelled chemical composition. The adopted stellar mass is color-coded.}
\label{Fig:Phi21_can}
\end{figure*}

Interesting trends can be noted when the Fourier parameters $R_{21}$ and $\Phi_{21}$ are plotted as a function of the pulsation period, as shown in Figures \ref{Fig:R21_can} and \ref{Fig:Phi21_can}. These plots report, for canonical models and the same masses and chemical compositions of previous figures (see labels), the trends of the predicted Fourier parameters, as inferred from the Gaia G-band light (left panels) and radial pulsation velocity (right panels) curves, as a function of the pulsation period.  In these plots the HP center, previously defined as the minimum in the pulsation amplitude can be identified with the maximum in the $R_{21}$ and $\Phi_{21}$ in the case of light curves and with the maximum in $R_{21}$ and in the slight flattening of $\Phi_{21}$ raising branch in the case of radial velocity curves.

A clear minimum in the case of $R_{21}$ and an evident discontinuity in the case of $\Phi_{21}$ are observed in the various panels for a period value that corresponds to the HP center, that significantly changes when varying the adopted stellar mass and chemical composition.

\begin{figure*}
\centering
\includegraphics[width=1.0\textwidth]{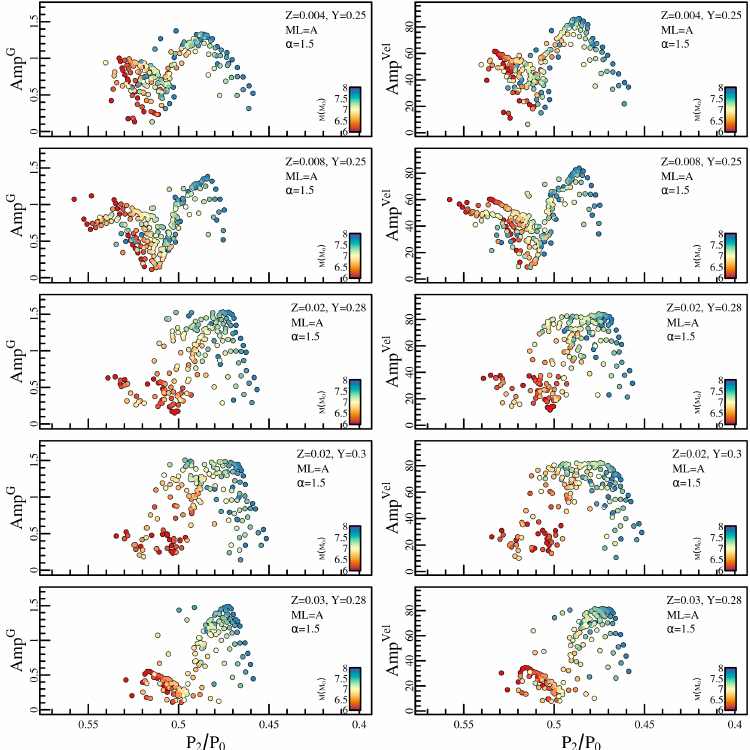}
\caption{The variation of the predicted Gaia G-band (left panels) and radial velocity (right panels) amplitudes as a function of the $P_{20}$ ratio for canonical models (case A) with the labelled chemical compositions. The mass value is color-coded in each panel.}
\label{Fig:amp_can-P20}
\end{figure*}



\subsection{Dependence of the central period of the progression on pulsation model inputs}
\begin{figure}
\centering
\includegraphics[trim={0 400 130 130},clip,width=0.5\textwidth]{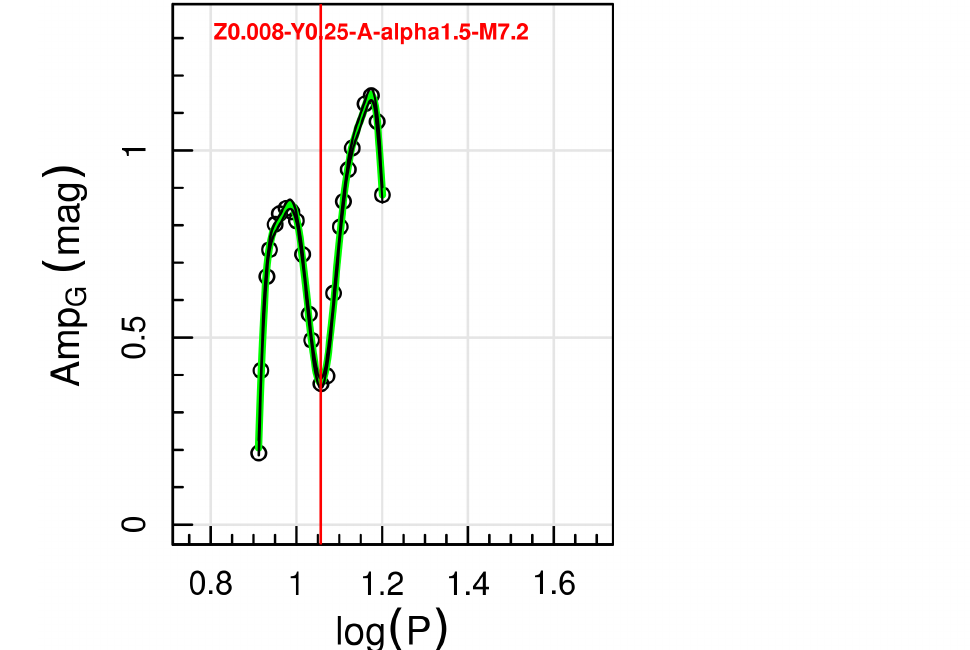}
\caption{Peak-to-peak amplitude as a function of the pulsation period, for models with Z=0.008, Y=0.25, $\alpha_{ml}=1.5$, canonical ML (case A) relation and $M=7.2M_\odot$, as fitted with the polynomial function (green line) adopted to estimate the HP central period (vertical red line).}
\label{fig:centralPeriodExample}
\end{figure}

To investigate in more detail the dependence of the HP central period on the various stellar parameters and model assumptions, we performed the following steps: i) the full sample of models was split into many sub-samples having fixed mass, elemental composition, $\alpha_{ml}$ and ML, and variable effective temperature (and period); ii) we visually inspected all these period (effective temperature) sequences, and selected only those crossing the centre of the progression, characterized by a central minimum in the peak-to-peak amplitude (see e.g. Figure ~\ref{fig:centralPeriodExample})  iii) following a procedure similar to that described in \citet{bhard15}, the central period of the progression, for the sequences defined at the previous point, was estimated by fitting, with a polynomial function\footnote{Different polynomial orders were tested and the best-fit one was set by requiring that the following residual variance was minimized:
\begin{equation}
    \sigma = \frac{Sr \left( m \right) }{n-m-1}
\end{equation}
where $Sr\left( m \right)$ is the sum of squared residuals, $n$ is the number of fitted points and $m$ is the order of the polynomial.}, the $G$ band peak-to-peak amplitude against the pulsation period, and selecting the period corresponding to the minimum amplitude. An example of this method is sketched in Figure ~\ref{fig:centralPeriodExample} where the sequence of models with $M$=$7.2M\odot$, Z=0.008, Y=0.25, canonical ML (case A) relation and $\alpha_{ml}$=1.5, is plotted in the plane of the $G$ band peak-to-peak amplitude and period, together with the polynomial function fitted to determine the central period of the progression.

\begin{figure}
\centering
\includegraphics[trim={30 150 0 110},clip,width=0.5\textwidth]{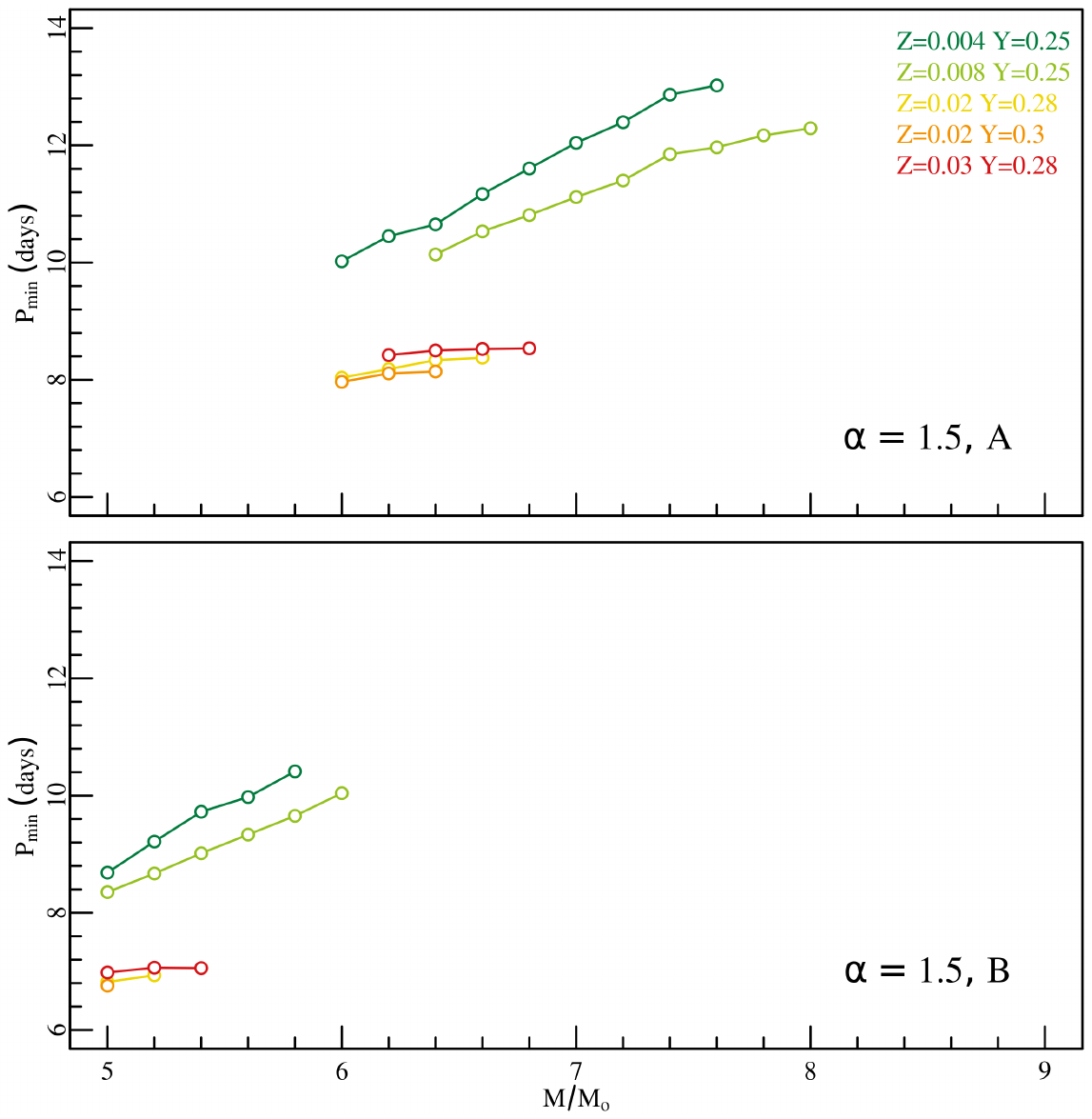}
\caption{The HP central period, computed by using the $Amp_G$ parameter, is plotted as a function of the model mass. The canonical (case A) and non-canonical (case B) models are plotted in the top and bottom panels, respectively. In each panel, different colors indicate different chemical compositions, as labeled in the legend.}
\label{fig:centralPeriodVsMass_ampG}
\end{figure}

Figure~\ref{fig:centralPeriodVsMass_ampG} shows the period at the HP center, estimated by using the G band peak-to-peak amplitude, as a function of the model mass. We notice that the period corresponding to the local minimum in the model amplitude steadily increases with the pulsation mass at the lower metal abundances, whereas a flatter behaviour is shown by solar and over-solar models. Moreover, we notice the above discussed decrease of the HP central period as the metallicity increases with a sort of saturation at the highest metal abundances, likely due to the smoother light curve morphology.

\section{Comparison with observations}

\begin{figure*}
\centering
\includegraphics[width=1.0\textwidth]{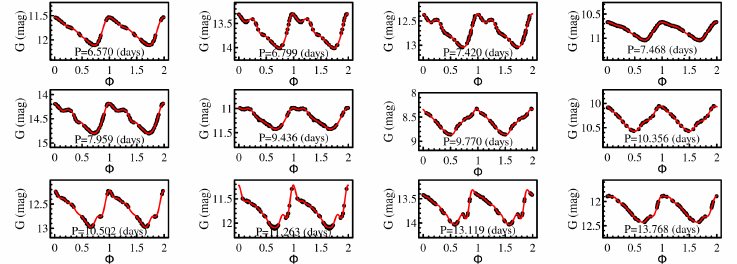}
\caption{The G band light curves of Cepheids within the Z1 sub-sample. The observational data points are plotted with black full symbols, while the Fourier model is represented by the red full line.}
\label{Fig:Z1-Dr3Sample}
\end{figure*}

\begin{figure*}
\centering
\includegraphics[page=1,width=1.0\textwidth]{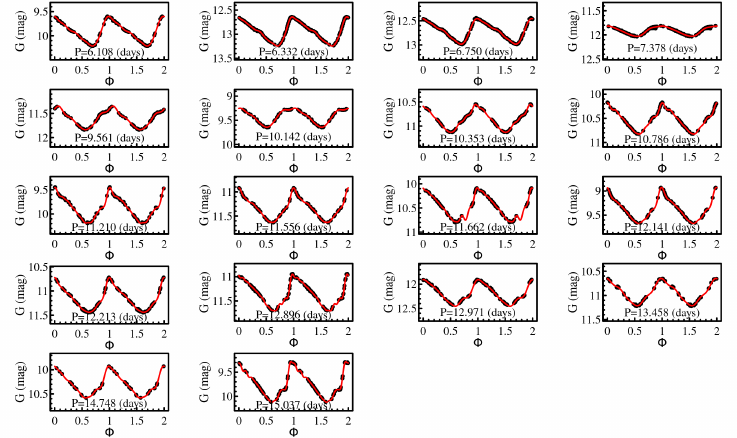}
\caption{The same as the previous figures but for the Z4 sub-sample.}
\label{Fig:Z4-Dr3Sample-pag1}
\end{figure*}

The light curves of Galactic bump Cepheids with the evidence of the HP phenomenon across a central period close to 8.2 days, as observed by the Gaia satellite and exemplified in the ESA Gaia Image of the Week of May 27, 2022  (https://www.cosmos.esa.int/web/gaia/iow\_20220527), can be an important benchmark for the above-presented pulsation models. However, metallicity differences among these pulsators might have an effect on the observed HP center, also considering the above-presented model predictions.
In order to build a sample of Classical Cepheids with metallicity spectroscopically determined, we focused our attention on the data provided within the C-MetaLL project \citep[see]{Ripepi2021, Trentin2023}, aimed at obtaining high-resolution spectroscopic data for Classical MW Cepheids, enlarging the sample of known objects towards the most metal-poor range ([Fe/H]$<-$0.4 dex). In particular, in the fourth paper of C-MetaLL collaboration \citep{Trentin2023b}, the Authors provide a list of 910 Classical Cepheids having accurate metallicities, including both the results from C-MetaLL projects and the literature. We searched this list for bump Cepheids in the  HP period range: $6 < P <16$ days, finding 261 objects for which we extracted the G band time series from the Gaia archive and modeled them with a truncated Fourier series. To select the best modeled time series, we focused on those sources fitted with more than 3 harmonics, and used the rms of residuals around the Fourier fit, together with the uniformity index (UI) introduced by \citet{Madore2005}. The latter parameter ranges between 0 and 1 and is a measure of the non-redundancy of the phase coverage and of the uniformity of the realized phase sampling. According to our tests, selecting only those sources with UI>0.95 allows us to avoid Fourier models with large spurious oscillations. Furthermore, together with the UI selection, we considered only those Fourier models with rms of residuals smaller than 0.01 mag. After this selection, the sample of observed Cepheids consists of 137 sources with good Fourier modeling. Taking into account the [Fe/H] values, we derived the corresponding global metallicities and divided the observational sample into four metallicity bins centered around Z = 0.01, Z=0.02, Z=0.03 and Z=0.04  (hereinafter Z1, Z2, Z3 and Z4). To take into account the metallicity error, we rejected sources with $\sigma_Z$ larger than $50\%$. Finally, the quoted metallicity bins Z1, Z2, Z3 and Z4, contain respectively 12, 41, 41 and 18 sources. The G band light curves for every sub-sample are plotted in  Figs. \ref{Fig:Z1-Dr3Sample}, \ref{Fig:Z2-Dr3Sample-pag1}, \ref{Fig:Z3-Dr3Sample-pag1} and \ref{Fig:Z4-Dr3Sample-pag1}.

\begin{figure*}
\centering
\includegraphics[page=1, width=1.0\textwidth]{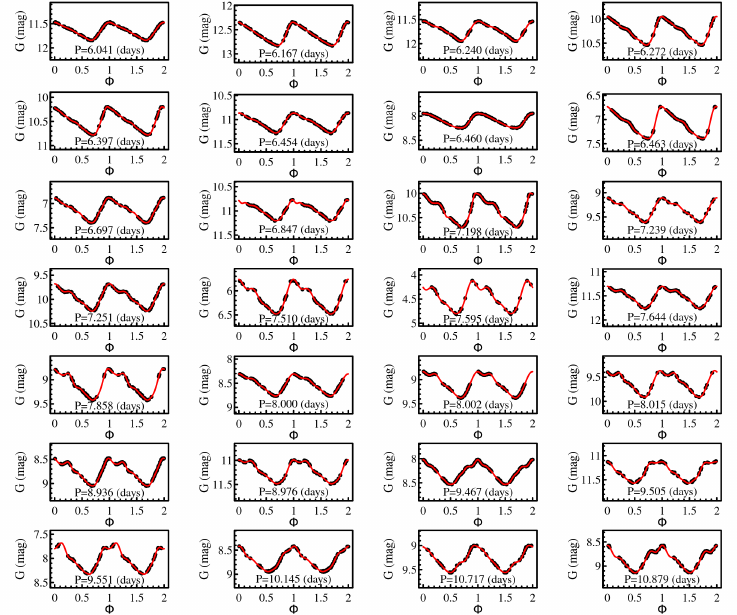}
\caption{The same as in Fig. \ref{Fig:Z1-Dr3Sample} but for the Z2 sub-sample.}
\label{Fig:Z2-Dr3Sample-pag1}
\end{figure*}

\begin{figure*}
\centering
\includegraphics[page=2, width=1.0\textwidth]{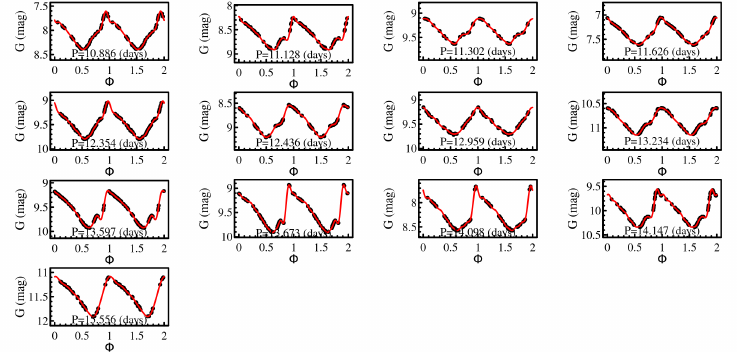}
\addtocounter{figure}{-1} 
\caption{continued}
\label{Fig:Z2-Dr3Sample-pag2}
\end{figure*}

\begin{figure*}
\centering
\includegraphics[page=1, width=1.0\textwidth]{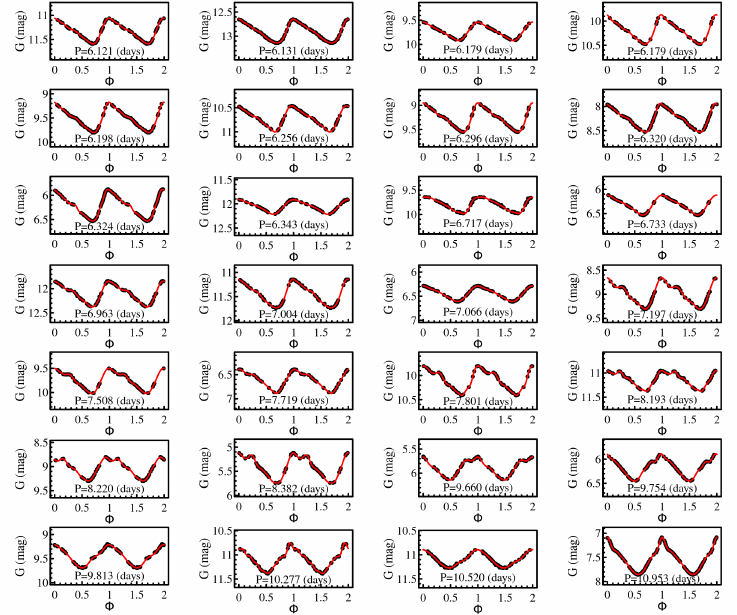}
\caption{The same as the previous figures but for the Z3 sub-sample.}
\label{Fig:Z3-Dr3Sample-pag1}
\end{figure*}

\begin{figure*}
\centering
\includegraphics[page=2, width=1.0\textwidth]{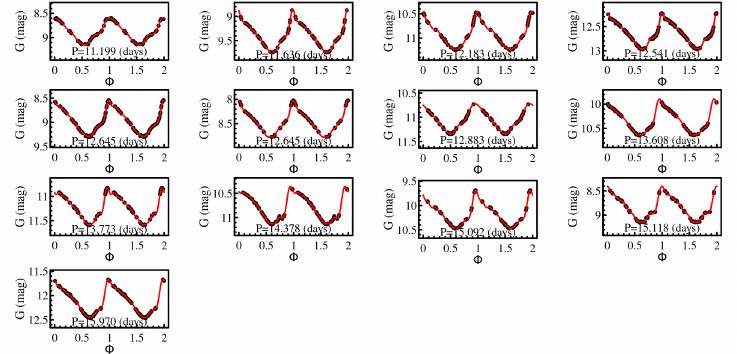}
\addtocounter{figure}{-1} 
\caption{continued}
\label{Fig:Z3-Dr3Sample-pag2}
\end{figure*}

Inspection of these figures suggests that, in qualitative agreement with theoretical indications, the period corresponding to the HP center tends to decrease as the metallicity increases. In particular, it decreases from $\sim$ 9.5 d to $\sim$ 7.5 d as the mean metallicity increases from $Z$=$0.01$ to $Z$=$0.04$. On the other hand, the presence of almost flat curves for different period values, within the same metallicity bin, might be ascribed to the already predicted and discussed dependence of the HP phenomenon on the stellar mass and ML relation.

\begin{figure}
\centering
\includegraphics[trim={0 140 130 0},clip,width=0.6\textwidth]{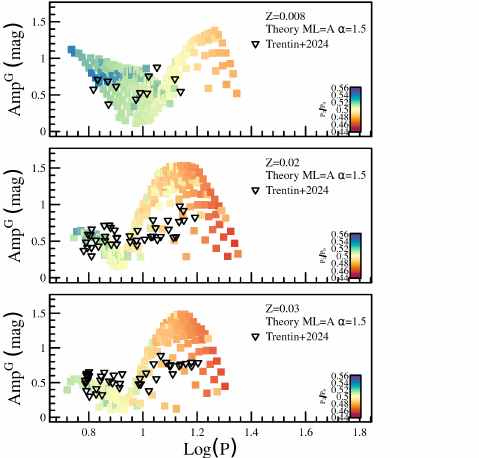}
\caption{Predicted G band light curves amplitudes compared with their observational counterparts in the period-amplitude plane. In each panel, the filled squares represent theoretical $\alpha_{ml}$=1.5 canonical values (case A), with color-coded $P_{20}$ values, while black symbols are the observational data. Different panels display results for different metal abundances: from top to bottom, the results for $Z$=$0.008$, $Z$=$0.02$ and $Z$=$0.03$, are compared with the sub-samples corresponding to the Z1, Z2 and Z3 metallicity bins, respectively.}
\label{Fig:Amp-logP-alpha1.5-mlA}
\end{figure}

\begin{figure}
\centering
\includegraphics[trim={0 140 130 0},clip,width=0.6\textwidth]{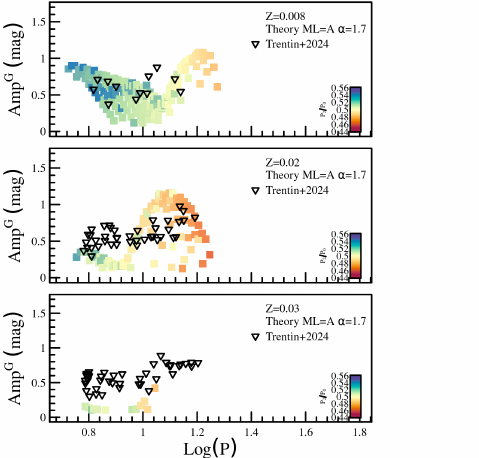}
\caption{The same as in Fig.\ref{Fig:Amp-logP-alpha1.5-mlA} but for $\alpha_{ml}$ = 1.7}
\label{Fig:Amp-logP-alpha1.7-mlA}
\end{figure}

\begin{figure}
\centering
\includegraphics[trim={0 140 130 0},clip,width=0.6\textwidth]{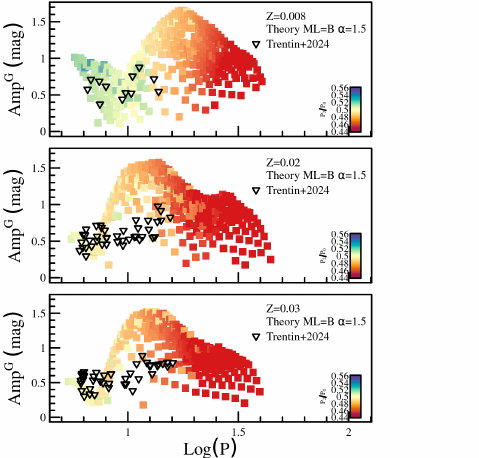}
\caption{The same as in Fig.\ref{Fig:Amp-logP-alpha1.5-mlA} but for a moderately non-canonical (case B) ML relation.}
\label{Fig:Amp-logP-alpha1.5-mlB}
\end{figure}

The pulsation G-band amplitudes of the light curves plotted in Figures \ref{Fig:Z1-Dr3Sample}, \ref{Fig:Z2-Dr3Sample-pag1} and \ref{Fig:Z3-Dr3Sample-pag1} are compared with model predictions in the period-amplitude plane in Figures \ref{Fig:Amp-logP-alpha1.5-mlA}, \ref{Fig:Amp-logP-alpha1.7-mlA} and \ref{Fig:Amp-logP-alpha1.5-mlB}, respectively.

We notice that canonical models (case A) with standard mixing length ($\alpha_{ml}$=1.5) roughly reproduce the amplitudes but poorly match their observed trend as the period increases. The agreement still worsens when increasing the efficiency of super-adiabatic convection.

In particular, when increasing the mixing length parameter from 1.5 to 1.7, smaller amplitudes than observed are predicted in the shorter period range, in particular at the higher metal abundances. 

On the other hand, the computed noncanonical models plotted in Fig. \ref{Fig:Amp-logP-alpha1.5-mlB} provide a slightly worse agreement than the canonical ones at solar and oversolar metallicity, even if for the lowest metal abundance bin, the minimum in the period-amplitude pattern seems to be better reproduced by these brighter models.

\begin{figure}
\centering
\includegraphics[trim={0 140 130 0},clip,width=0.6\textwidth]{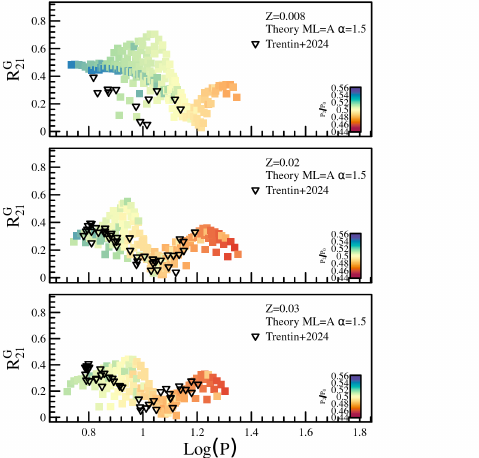}
\caption{The same as fig.\ref{Fig:Amp-logP-alpha1.5-mlA} but for the G band Fourier parameter $R_{21}$.}
\label{Fig:R21-logP-alpha1.5-mlA}
\end{figure}

\begin{figure}
\centering
\includegraphics[trim={0 140 130 0},clip,width=0.6\textwidth]{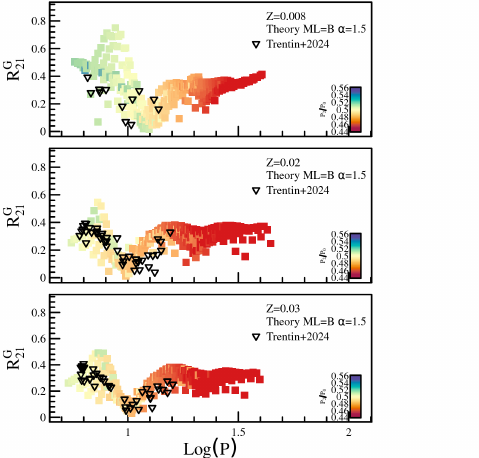}
\caption{The same as in Fig. \ref{Fig:R21-logP-alpha1.5-mlA} but for a moderately noncanonical (case B) ML relation.}
\label{Fig:R21-logP-alpha1.5-mlB}
\end{figure}

Similar comparisons are performed for the Fourier parameter $R_{21}$ in Figures \ref{Fig:R21-logP-alpha1.5-mlA} and \ref{Fig:R21-logP-alpha1.5-mlB} for the canonical and noncanonical cases, respectively. 

We notice that the predicted $R_{21}$ pattern satisfactorily reproduces the observed trend as a function of the period, in particular at solar and over-solar metallicity, with a slightly better agreement in the case of non-canonical models

\begin{figure}
\centering
\includegraphics[trim={0 140 130 0},clip,width=0.6\textwidth]{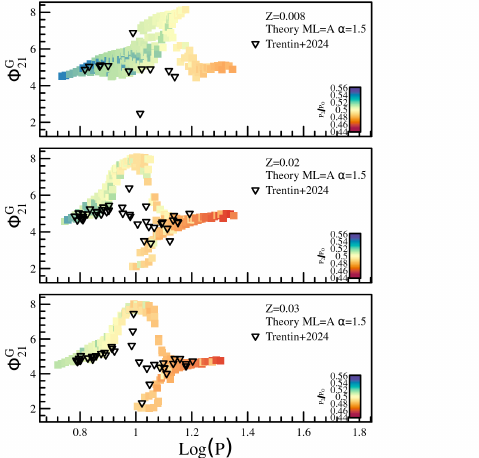}
\caption{The same as in Fig.\ref{Fig:Amp-logP-alpha1.5-mlA} but for the G band Fourier parameter $\Phi_{21}$.}
\label{Fig:Phi21-logP-alpha1.5-mlA}
\end{figure}

\begin{figure}
\centering
\includegraphics[trim={0 140 130 0},clip,width=0.6\textwidth]{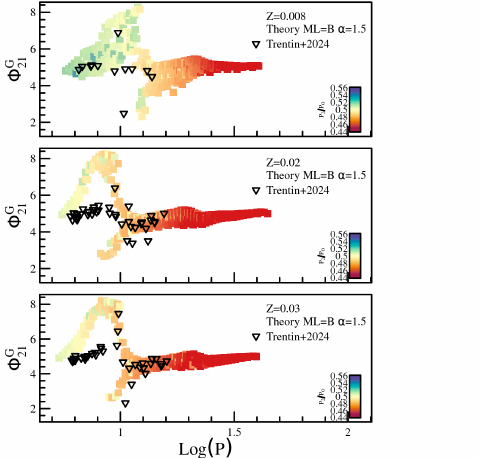}
\caption{The same as in Fig.\ref{Fig:Phi21-logP-alpha1.5-mlA} but moderately noncanonical models (case B).}
\label{Fig:Phi21-logP-alpha1.5-mlB}
\end{figure}

Figures \ref{Fig:Phi21-logP-alpha1.5-mlA} and \ref{Fig:Phi21-logP-alpha1.5-mlB}
display the trend of the $\Phi_{21}$ Fourier parameter as a function of the pulsation period for the same metallicity ranges as in previous figures, assuming a canonical and a non-canonical ML relation, respectively. 

In these plots, the main observed features are roughly predicted by models with a slightly better agreement for $Z$=$0.02$ and $Z$=$0.03$ canonical and $Z$=$0.008$ non-canonical models, even if we notice a main discrepancy at short periods, with the theoretical $\Phi_{21}$ Fourier parameters
 systematically higher than the observed ones.
 
A larger number of observed Cepheids with known metal abundances 
should be included to test the predicted metallicity dependence of the HP phenomenon, as well as the effect of the ML relation (and stellar mass).

\section{Conclusions}

A fine grid of non-linear convective pulsation models of bump Cepheids has been computed in order to investigate the HP phenomenon shown by their light and radial pulsation velocity curves. 
Several trends of the HP center have been investigated through inspection of the light and radial velocity curves morphology as well as of the period-amplitude, period-$R_{21}$ and period-$\Phi_{21}$ plots.
\begin{itemize}
    \item A variation in the efficiency of super-adiabatic convection only affects the pulsation amplitudes but does not change the center of the HP.
    \item At fixed chemical composition and ML relation, the HP center period moves towards longer values as the stellar mass increases, but with a flattening of this trend for the highest metal abundances.
    \item The same effects are produced when assuming a brighter ML relation.
    \item At fixed ML relation, and rather independently of the mass value, the HP center period moves towards shorter values as the metallicity increases from $Z$=0.004 to solar and over-solar abundances, with a smaller effect of the helium abundance at least at solar metallicity.
 \item  The evidence that models with similar P2/P0 have similar light and velocity curves seems to support the crucial role of resonance in shaping the Hertzsprung Progression.
\end{itemize}

To test these model predictions, we selected a sample of 112 observed bump Cepheids in the Gaia database with metal abundances from the C-MetaLL survey \citep{Trentin2023b}, in the  HP period range: $6 < P <16$ days. 
Four metallicity bins have been identified to investigate the HP phenomenon of observed light curves. The HP center period shows a trend with metallicity similar to the predicted one, decreasing from $\sim$ 9.5 d to $\sim$ 7.5 d as the mean metallicity increases from $Z$=$0.01$ to $Z$=$0.04$. 
From the selected empirical light curves we also derived the pulsation amplitudes and the Fourier parameters that were compared with model predictions in the period-amplitude, period-$R_{21}$ and period-$\Phi_{21}$ planes. 

The main observed features in these plots are satisfactorily reproduced by models but additional observations are needed in order to draw quantitative conclusions both on the dependence of the HP center on metallicity and on the effect of the ML relation and the stellar mass.

\section*{Acknowledgements}

We thank the anonymous referee for useful comments and suggestions that contributed to improve the quality of the paper. This project has received funding from the European Union's Horizon 2020 research and innovation program under the Marie Sklodowska- Curie grant agreement No. 886298.  We acknowledge the financial support from ASI-Gaia (“Missione Gaia Partecipazione italiana al DPAC – Operazioni e Attività di Analisi dati”). 
This research was also supported by the International Space Science Institute (ISSI) in Bern, through ISSI International Team project SHoT: The Stellar Path to the Ho Tension in the Gaia, TESS, LSST and JWST Era.
GDS thanks the support from Istituto Nazionale di Fisica Nucleare (INFN), Naples section-Specific Initiatives QGSKY, and Moonlight2. 

Data availability statement:The data underlying this article are available in the article and in its online supplementary material.



\bibliographystyle{mnras}
\bibliography{mar} 

\appendix

\section{Stellar parameters of computed pulsation models}\label{app:the-model-parameters}
\onecolumn
\begin{longtable}{cccccccc}
\caption{\label{param_HP_models} The adopted intrinsic stellar parameters.
Columns from 1 to 8 list the metallicity, the helium content, the stellar mass, the luminosity level, the mixing length parameter, the ML relation label (A - canonical, B - noncanonical), the FBE, and the FRE effective temperatures.}\\
\hline\hline
Z & Y & \msun & \lsun & $\alpha_{ml}$ & ML & FBE & FRE \\
\hline
\endfirsthead
\caption{continued.}\\
\hline\hline
Z & Y & \msun & \lsun & $\alpha_{ml}$ & ML & FBE & FRE \\
\hline
\endhead
\hline
0.004&0.25&5.0&3.44&1.5&B&5925&5025\\
0.004&0.25&5.2&3.50&1.5&B&6025&4975\\
0.004&0.25&5.4&3.55&1.5&B&6025&4925\\
0.004&0.25&5.6&3.60&1.5&B&6025&4875\\
0.004&0.25&5.8&3.65&1.5&B&5975&4825\\
0.004&0.25&6.0&3.5&1.5&A&5825&5075\\
0.004&0.25&6.0&3.5&1.7&A&5925&5275\\
0.004&0.25&6.0&3.7&1.5&B&5975&4775\\
0.004&0.25&6.0&3.7&1.7&B&5925&5025\\
0.004&0.25&6.2&3.55&1.5&A&5825&5075\\
0.004&0.25&6.2&3.55&1.7&A&5975&5225\\
0.004&0.25&6.2&3.75&1.5&B&5925&4725\\
0.004&0.25&6.2&3.75&1.7&B&5875&4975\\
0.004&0.25&6.4&3.6&1.5&A&5875&4975\\
0.004&0.25&6.4&3.6&1.7&A&6025&5175\\
0.004&0.25&6.4&3.8&1.5&B&5875&4675\\
0.004&0.25&6.4&3.8&1.7&B&5875&4925\\
0.004&0.25&6.6&3.64&1.5&A&5875&4925\\
0.004&0.25&6.6&3.64&1.7&A&5975&5125\\
0.004&0.25&6.6&3.84&1.5&B&5875&4625\\
0.004&0.25&6.6&3.84&1.7&B&5825&4875\\
0.004&0.25&6.8&3.69&1.5&A&5925&4925\\
0.004&0.25&6.8&3.69&1.7&A&5975&5125\\
0.004&0.25&6.8&3.89&1.5&B&5875&4575\\
0.004&0.25&6.8&3.89&1.7&B&5825&4825\\
0.004&0.25&7.0&3.73&1.5&A&5925&4875\\
0.004&0.25&7.0&3.73&1.7&A&5975&5075\\
0.004&0.25&7.0&3.93&1.5&B&5775&4525\\
0.004&0.25&7.0&3.93&1.7&B&5775&4825\\
0.004&0.25&7.2&3.77&1.5&A&5925&4825\\
0.004&0.25&7.2&3.77&1.7&A&5925&5025\\
0.004&0.25&7.2&3.97&1.5&B&5775&4525\\
0.004&0.25&7.2&3.97&1.7&B&5775&4775\\
0.004&0.25&7.4&3.81&1.5&A&5925&4825\\
0.004&0.25&7.4&3.81&1.7&A&5925&5025\\
0.004&0.25&7.4&4.01&1.5&B&5775&4425\\
0.004&0.25&7.4&4.01&1.7&B&5775&4725\\
0.004&0.25&7.6&3.85&1.5&A&5875&4775\\
0.004&0.25&7.6&3.85&1.7&A&5875&4975\\
0.004&0.25&7.6&4.05&1.5&B&5775&4475\\
0.004&0.25&7.6&4.05&1.7&B&5725&4675\\
0.004&0.25&7.8&3.89&1.5&A&5875&4725\\
0.004&0.25&7.8&3.89&1.7&A&5875&4975\\
0.004&0.25&7.8&4.09&1.5&B&5725&4475\\
0.004&0.25&7.8&4.09&1.7&B&5725&4625\\
0.004&0.25&8.0&3.92&1.5&A&5825&4725\\
0.004&0.25&8.0&3.92&1.7&A&5825&4925\\
0.004&0.25&8.0&4.12&1.5&B&5725&4475\\
0.004&0.25&8.0&4.12&1.7&B&5725&4625\\
0.008&0.25&5.0&3.34&1.5&B&5825&5025\\
0.008&0.25&5.2&3.39&1.5&B&5875&4975\\
0.008&0.25&5.4&3.45&1.5&B&5925&4925\\
0.008&0.25&5.6&3.50&1.5&B&5975&4875\\
0.008&0.25&5.8&3.55&1.5&B&5975&4825\\
0.008&0.25&6.0&3.4&1.5&A&5925&5125\\
0.008&0.25&6.0&3.4&1.7&A&5975&5275\\
0.008&0.25&6.0&3.6&1.5&B&5925&4775\\
0.008&0.25&6.0&3.6&1.7&B&5925&5025\\
0.008&0.25&6.2&3.45&1.5&A&5775&5075\\
0.008&0.25&6.2&3.45&1.7&A&5975&5225\\
0.008&0.25&6.2&3.65&1.5&B&5875&4725\\
0.008&0.25&6.2&3.65&1.7&B&5875&4975\\
0.008&0.25&6.4&3.49&1.5&A&5775&5025\\
0.008&0.25&6.4&3.49&1.7&A&5925&5225\\
0.008&0.25&6.4&3.69&1.5&B&5875&4675\\
0.008&0.25&6.4&3.69&1.7&B&5825&4925\\
0.008&0.25&6.6&3.54&1.5&A&5775&4975\\
0.008&0.25&6.6&3.54&1.7&A&5925&5175\\
0.008&0.25&6.6&3.74&1.5&B&5825&4625\\
0.008&0.25&6.6&3.74&1.7&B&5775&4875\\
0.008&0.25&6.8&3.58&1.5&A&5825&4925\\
0.008&0.25&6.8&3.58&1.7&A&5925&5125\\
0.008&0.25&6.8&3.78&1.5&B&5825&4575\\
0.008&0.25&6.8&3.78&1.7&B&5775&4825\\
0.008&0.25&7.0&3.63&1.5&A&5825&4875\\
0.008&0.25&7.0&3.63&1.7&A&5875&5075\\
0.008&0.25&7.0&3.83&1.5&B&5775&4575\\
0.008&0.25&7.0&3.83&1.7&B&5725&4825\\
0.008&0.25&7.2&3.67&1.5&A&5925&4825\\
0.008&0.25&7.2&3.67&1.7&A&5875&5025\\
0.008&0.25&7.2&3.87&1.5&B&5725&4525\\
0.008&0.25&7.2&3.87&1.7&B&5725&4775\\
0.008&0.25&7.4&3.71&1.5&A&5875&4775\\
0.008&0.25&7.4&3.71&1.7&A&5825&5025\\
0.008&0.25&7.4&3.91&1.5&B&5725&4475\\
0.008&0.25&7.4&3.91&1.7&B&5725&4725\\
0.008&0.25&7.6&3.74&1.5&A&5875&4775\\
0.008&0.25&7.6&3.74&1.7&A&5775&4975\\
0.008&0.25&7.6&3.94&1.5&B&5675&4475\\
0.008&0.25&7.6&3.94&1.7&B&5675&4675\\
0.008&0.25&7.8&3.78&1.5&A&5825&4725\\
0.008&0.25&7.8&3.78&1.7&A&5775&4975\\
0.008&0.25&7.8&3.98&1.5&B&5675&4425\\
0.008&0.25&7.8&3.98&1.7&B&5675&4625\\
0.008&0.25&8.0&3.82&1.5&A&5775&4675\\
0.008&0.25&8.0&3.82&1.7&A&5775&4925\\
0.008&0.25&8.0&4.02&1.5&B&5625&4425\\
0.008&0.25&8.0&4.02&1.7&B&5625&4625\\
0.02&0.28&5.0&3.27&1.5&B&5725&4925\\
0.02&0.28&5.2&3.32&1.5&B&5675&4875\\
0.02&0.28&5.4&3.38&1.5&B&5625&4825\\
0.02&0.28&5.6&3.43&1.5&B&5625&4725\\
0.02&0.28&5.8&3.48&1.5&B&5625&4675\\
0.02&0.28&6.0&3.33&1.5&A&5725&4975\\
0.02&0.28&6.0&3.33&1.7&A&5675&5225\\
0.02&0.28&6.0&3.53&1.5&B&5575&4625\\
0.02&0.28&6.0&3.53&1.7&B&5475&4925\\
0.02&0.28&6.2&3.38&1.5&A&5675&4875\\
0.02&0.28&6.2&3.38&1.7&A&5625&5225\\
0.02&0.28&6.2&3.58&1.5&B&5525&4575\\
0.02&0.28&6.2&3.58&1.7&B&5425&4875\\
0.02&0.28&6.4&3.43&1.5&A&5625&4825\\
0.02&0.28&6.4&3.43&1.7&A&5625&5125\\
0.02&0.28&6.4&3.63&1.5&B&5475&4525\\
0.02&0.28&6.4&3.63&1.7&B&5375&4825\\
0.02&0.28&6.6&3.47&1.5&A&5625&4825\\
0.02&0.28&6.6&3.47&1.7&A&5525&5125\\
0.02&0.28&6.6&3.67&1.5&B&5425&4475\\
0.02&0.28&6.6&3.67&1.7&B&5325&4775\\
0.02&0.28&6.8&3.51&1.5&A&5575&4775\\
0.02&0.28&6.8&3.51&1.7&A&5475&5025\\
0.02&0.28&6.8&3.71&1.5&B&5425&4475\\
0.02&0.28&6.8&3.71&1.7&B&5275&4725\\
0.02&0.28&7.0&3.56&1.5&A&5525&4725\\
0.02&0.28&7.0&3.56&1.7&A&5375&5025\\
0.02&0.28&7.0&3.76&1.5&B&5375&4425\\
0.02&0.28&7.0&3.76&1.7&B&5225&4725\\
0.02&0.28&7.2&3.6&1.5&A&5475&4725\\
0.02&0.28&7.2&3.6&1.7&A&5375&4975\\
0.02&0.28&7.2&3.8&1.5&B&5325&4325\\
0.02&0.28&7.2&3.8&1.7&B&5225&4625\\
0.02&0.28&7.4&3.64&1.5&A&5475&4675\\
0.02&0.28&7.4&3.64&1.7&A&5375&4925\\
0.02&0.28&7.4&3.84&1.5&B&5275&4275\\
0.02&0.28&7.4&3.84&1.7&B&5175&4625\\
0.02&0.28&7.6&3.68&1.5&A&5475&4625\\
0.02&0.28&7.6&3.68&1.7&A&5325&4925\\
0.02&0.28&7.6&3.88&1.5&B&5275&4275\\
0.02&0.28&7.6&3.88&1.7&B&5125&4575\\
0.02&0.28&7.8&3.71&1.5&A&5425&4575\\
0.02&0.28&7.8&3.71&1.7&A&5325&4875\\
0.02&0.28&7.8&3.91&1.5&B&5225&4225\\
0.02&0.28&7.8&3.91&1.7&B&5075&4525\\
0.02&0.28&8.0&3.75&1.5&A&5375&4525\\
0.02&0.28&8.0&3.75&1.7&A&5275&4875\\
0.02&0.28&8.0&3.95&1.5&B&5225&4175\\
0.02&0.28&8.0&3.95&1.7&B&5075&4475\\
0.02&0.3&5.0&3.31&1.5&B&5725&4925\\
0.02&0.3&5.2&3.36&1.5&B&5675&4875\\
0.02&0.3&5.4&3.42&1.5&B&5675&4775\\
0.02&0.3&5.6&3.47&1.5&B&5625&4725\\
0.02&0.3&5.8&3.52&1.5&B&5575&4725\\
0.02&0.3&6.0&3.37&1.5&A&5725&4925\\
0.02&0.3&6.0&3.37&1.7&A&5575&5325\\
0.02&0.3&6.0&3.57&1.5&B&5525&4625\\
0.02&0.3&6.0&3.57&1.7&B&5375&4925\\
0.02&0.3&6.2&3.42&1.5&A&5675&4925\\
0.02&0.3&6.2&3.42&1.7&A&5575&5225\\
0.02&0.3&6.2&3.62&1.5&B&5475&4575\\
0.02&0.3&6.2&3.62&1.7&B&5325&4875\\
0.02&0.3&6.4&3.47&1.5&A&5625&4875\\
0.02&0.3&6.4&3.47&1.7&A&5425&5175\\
0.02&0.3&6.4&3.67&1.5&B&5475&4525\\
0.02&0.3&6.4&3.67&1.7&B&5275&4825\\
0.02&0.3&6.6&3.51&1.5&A&5575&4825\\
0.02&0.3&6.6&3.51&1.7&A&5275&5125\\
0.02&0.3&6.6&3.71&1.5&B&5425&4475\\
0.02&0.3&6.6&3.71&1.7&B&5225&4825\\
0.02&0.3&6.8&3.56&1.5&A&5575&4775\\
0.02&0.3&6.8&3.56&1.7&A&5425&5225\\
0.02&0.3&6.8&3.76&1.5&B&5375&4425\\
0.02&0.3&6.8&3.76&1.7&B&5225&4775\\
0.02&0.3&7.0&3.6&1.5&A&5475&4725\\
0.02&0.3&7.0&3.6&1.7&A&5375&5025\\
0.02&0.3&7.0&3.8&1.5&B&5325&4375\\
0.02&0.3&7.0&3.8&1.7&B&5175&4675\\
0.02&0.3&7.2&3.64&1.5&A&5425&4675\\
0.02&0.3&7.2&3.64&1.7&A&5325&5025\\
0.02&0.3&7.2&3.84&1.5&B&5275&4325\\
0.02&0.3&7.2&3.84&1.7&B&5125&4675\\
0.02&0.3&7.4&3.68&1.5&A&5425&4675\\
0.02&0.3&7.4&3.68&1.7&A&5275&4975\\
0.02&0.3&7.4&3.88&1.5&B&5275&4275\\
0.02&0.3&7.4&3.88&1.7&B&5075&4625\\
0.02&0.3&7.6&3.72&1.5&A&5425&4575\\
0.02&0.3&7.6&3.72&1.7&A&5225&4925\\
0.02&0.3&7.6&3.92&1.5&B&5225&4225\\
0.02&0.3&7.6&3.92&1.7&B&5025&4575\\
0.02&0.3&7.8&3.76&1.5&A&5375&4525\\
0.02&0.3&7.8&3.76&1.7&A&5175&4925\\
0.02&0.3&7.8&3.96&1.5&B&5175&4175\\
0.02&0.3&7.8&3.96&1.7&B&5125&4525\\
0.02&0.3&8.0&3.79&1.5&A&5375&4525\\
0.02&0.3&8.0&3.79&1.7&A&5175&4925\\
0.02&0.3&8.0&3.99&1.5&B&5175&4125\\
0.02&0.3&8.0&3.99&1.7&B&5125&4475\\
0.03&0.28&5.0&3.21&1.5&B&5575&4875\\
0.03&0.28&5.2&3.26&1.5&B&5525&4825\\
0.03&0.28&5.4&3.32&1.5&B&5525&4775\\
0.03&0.28&5.6&3.37&1.5&B&5475&4725\\
0.03&0.28&5.8&3.42&1.5&B&5425&4675\\
0.03&0.28&6.0&3.27&1.5&A&5625&4925\\
0.03&0.28&6.0&3.27&1.7&A&5375&5225\\
0.03&0.28&6.0&3.47&1.5&B&5425&4625\\
0.03&0.28&6.0&3.47&1.7&B&5225&4925\\
0.03&0.28&6.2&3.32&1.5&A&5525&4925\\
0.03&0.28&6.2&3.32&1.7&A&5375&5225\\
0.03&0.28&6.2&3.52&1.5&B&5375&4575\\
0.03&0.28&6.2&3.52&1.7&B&5225&4875\\
0.03&0.28&6.4&3.37&1.5&A&5475&4825\\
0.03&0.28&6.4&3.37&1.7&A&5375&5175\\
0.03&0.28&6.4&3.57&1.5&B&5325&4525\\
0.03&0.28&6.4&3.57&1.7&B&5125&4825\\
0.03&0.28&6.6&3.41&1.5&A&5525&4775\\
0.03&0.28&6.6&3.41&1.7&A&5325&5175\\
0.03&0.28&6.6&3.61&1.5&B&5325&4475\\
0.03&0.28&6.6&3.61&1.7&B&5125&4825\\
0.03&0.28&6.8&3.45&1.5&A&5475&4725\\
0.03&0.28&6.8&3.45&1.7&A&5225&5125\\
0.03&0.28&6.8&3.65&1.5&B&5275&4425\\
0.03&0.28&6.8&3.65&1.7&B&5075&4725\\
0.03&0.28&7.0&3.5&1.5&A&5425&4675\\
0.03&0.28&7.0&3.5&1.7&A&5225&4975\\
0.03&0.28&7.0&3.7&1.5&B&5225&4375\\
0.03&0.28&7.0&3.7&1.7&B&5025&4725\\
0.03&0.28&7.2&3.54&1.5&A&5375&4675\\
0.03&0.28&7.2&3.54&1.7&A&5175&5025\\
0.03&0.28&7.2&3.74&1.5&B&5175&4325\\
0.03&0.28&7.2&3.74&1.7&B&4975&4675\\
0.03&0.28&7.4&3.58&1.5&A&5375&4625\\
0.03&0.28&7.4&3.58&1.7&A&4975&4925\\
0.03&0.28&7.4&3.78&1.5&B&5125&4275\\
0.03&0.28&7.4&3.78&1.7&B&4925&4575\\
0.03&0.28&7.6&3.62&1.5&A&5325&4575\\
0.03&0.28&7.6&3.62&1.7&A&5075&4975\\
0.03&0.28&7.6&3.82&1.5&B&5125&4225\\
0.03&0.28&7.6&3.82&1.7&B&4875&4575\\
0.03&0.28&7.8&3.65&1.5&A&5275&4525\\
0.03&0.28&7.8&3.85&1.5&B&5075&4175\\
0.03&0.28&7.8&3.85&1.7&B&4875&4525\\
0.03&0.28&8.0&3.69&1.5&A&5275&4525\\
0.03&0.28&8.0&3.69&1.7&A&5075&4925\\
0.03&0.28&8.0&3.89&1.5&B&5025&4175\\
0.03&0.28&8.0&3.89&1.7&B&4825&4475\\
\hline\hline
\end{longtable}

\twocolumn

\bsp	
\label{lastpage}
\end{document}